\documentclass{aa}

\usepackage{algorithm}
\usepackage[end,]{algpseudocode}
\usepackage{amssymb}
\usepackage{amsmath}
\usepackage{caption}
\usepackage{datetime}
\usepackage{graphicx}
\usepackage{hyperref}
\usepackage{layouts}
\usepackage{mathtools}
\usepackage{natbib}

\usepackage{placeins}
\usepackage{setspace}
\usepackage{siunitx}
\usepackage{subcaption}

\usepackage{txfonts}
\usepackage{xcolor}
\usepackage{svg}

\algnewcommand{\LineComment}[1]{\Statex \(\triangleright\) #1}
\DeclareSIUnit \Jy {Jy}
\DeclareSIUnit \px {px}
\DeclareSIUnit \beam {beam}
\DeclareMathOperator{\relu}{ReLU}
\newcommand{\printHsizeInCm}{
  \strip@pt\dimexpr\hsize*65536/1cm\relax
}

\usepackage{xparse}
\makeatletter
\NewDocumentCommand{\LeftComment}{s m}{
  \Statex \IfBooleanF{#1}{\hspace*{\ALG@thistlm}}\(\triangleright\) #2}
\makeatother
\begin{document}

   \title{Generating radio continuum survey maps of arbitrary size with latent diffusion models}

   \author{
        T. Vičánek Martínez \inst{1}
        \and M. Brüggen \inst{1}
    }
    
   \institute{
        Hamburger Sternwarte, Universität Hamburg, Gojenbergsweg 112, 21029 Hamburg, Germany
   }
   \date{}

  \abstract
  {Radio surveys provide exponentially increasing amounts of observational data, which requires the development of novel data reduction and analysis methods. To this end, realistic simulations of observational data provide controlled environments for testing and developing new methods. Diffusion models excel at synthesizing realistic image data, but are limited by size constraints and high computational demands.}
  {We implemented a latent diffusion model (LDM) to synthesize realistic radio survey map cutouts. Further, we developed a technique for sampling radio survey maps of arbitrary size with the LDM. This is done by sampling in multiple sequential steps, whereby consistency is provided through context vectors and image inpainting.}
  {We trained our model on observations from the LOFAR telescope obtained from the third data release of the LOFAR Two-Metre Sky Survey. We implemented a LDM composed of a vector-quantized variational autoencoder and a diffusion model operating on the resulting latent representations. We added information on source locations, brightness and sizes by extracting parameters from the public source catalog and encoding them into a vector format passed as context to the LDM. We trained the model for image inpainting to facilitate continuous sequential sampling.}
  {We were able to generate realistic cutouts of 512 pixel side length, corresponding to $\sim\!\ang{;5.7;}$, with precise control over source locations and brightness. Control over source extension was achieved at lower precision. Sampling of arbitrary-sized maps was made possible, showing realistic results with seamless combination of individual samples and limited only by occasional sampling artifacts.}
  {}


   \maketitle

\section{Introduction} \label{sec:Introduction}

Following the continuously improving capacities of radio astronomical instruments experienced over the past decades, radio sky surveys provide observational data at exponentially increasing volumes \citep{Smith&Geach23}. Given the richness of such observations that cover wide angular scales at high resolution and sensitivities, these advances open substantial opportunities for discoveries in diverse areas, including the evolution of galaxies and active galactic nuclei \citep[e.g.][]{Best+2014,Wilman+2008}, large-scale structure and cosmology \citep[e.g.][]{Schwarz+2015}, as well as the study of magnetic fields and the intergalactic medium \citep[e.g.][]{Cassano+2010}.\\
As data taking capabilities advance, the primary bottleneck shifts from acquiring observations to efficiently processing, analyzing and interpreting them. Handling such vast amounts of data requires sophisticated computational methods that both scale with data volume and support automated execution. As conventional analysis of radio astronomical data often involves human supervision and expert knowledge, future discoveries inevitably entail a demand for novel approaches to execute tasks like source detection and classification. In this context, simulations play an increasingly important role: they support the development and validation of data analysis pipelines and provide controlled environments to test methods developed for the next generation of radio surveys.\\
Accurately synthesizing realistic radio observational data is however no straightforward task. The main complications arise from the intrinsic complexity of both the astrophysical sources and the measurement process. Radio source morphologies span a wide range of scales and structures - from compact point-like emitters to extended objects with a variety of shapes such as radio jets, lobes, and diffuse emission - a mix for which no single physical model or simulation framework can adequately capture the full diversity. In addition, the response of radio interferometers is governed by a combination of instrumental effects, calibration errors, and atmospheric influences, leading to nontrivial noise properties, systematic signals, and imaging artifacts that are challenging to model accurately. Finally, radio sources are not uniformly distributed across the sky but exhibit spatial correlations driven by large-scale structure and influenced by survey selection effects, which must be faithfully reproduced to generate statistically representative observational data. While existing approaches for simulating radio observations, such as the SKA Science Data Challenges, OSKAR, and RASCIL, offer detailed models of instrumental and telescope effects \citep{Bonaldi+2021,Hartley+2023,Dulwich+2009,Cornwell+2025}, their intrinsic source morphology models are restricted to simple Gaussian shapes or a limited library of observed sources, and thus they do not fully capture the statistical diversity of extended radio emission.\\
Luckily however, the plethora of newly gathered observational data does not only pose new challenges, but also creates novel opportunities at the same time. Recently, deep learning–based generative approaches have emerged as powerful tools for synthesizing image data. In particular, diffusion models (DMs) have demonstrated remarkable performance in high-fidelity generation of natural images \citep{Dhariwal+21, Karras+22, Karras+23}. Considering how the performance of machine learning models scales with the amount and variety of data they are trained on, DMs therefore qualify as promising candidates for potential use in applications related to astronomical image data. In that context, this class of generative models has been employed for generation of single-source images \citep[e.g.][]{Smith+2022, VicanekMartinez24,Ma+2026,Potevineau+2026, Scognamiglio+2025}, image deconvolution and super-resolution \citep{Reddy+24,Shan+2025,Spagnoletti+2024}, radio-astronomical image reconstruction \citep{Wang+2023, Drozdova+2024}, and for removal of noise and artifacts \citep{WaldmannRocchetto2023, Nicolaas+2025}.\\
DMs work by training a neural network to remove artificial noise from training images. Once trained, this network can produce novel images by iteratively removing small amounts of noise in several steps, starting from an initial seed image of pure random noise. While successful, their application is limited by practical constraints \citep[see e.g.][]{Chen+2025}. Due to the complexity of model architectures and the variability of data distributions in high-dimensional image space, DMs require substantial computational resources for training. The same issue also occurs during inference, given the sequential nature of the sampling procedure that requires multiple evaluations of the trained model. Furthermore, DMs are generally restricted to fixed image sizes, to which the aforementioned computational demands impose an upper limit. These issues present a challenge for the deployment of DMs on the large, high-resolution datasets characteristic of modern radio surveys. Dealing with these methodological constraints is essential to fully exploit DMs for the generation of synthetic survey observations with next-generation radio telescopes.\\
The aim of this work is to address these limitations by implementing a generative machine learning model designed for high-quality generation of large-size synthetic radio survey maps. To this end, we employ a variant of DMs referred to as latent diffusion models (LDMs). First introduced in \cite{Rombach+2021}, this method works by combining DMs with another class of generative models called variational autoencoders (VAEs). These consist of an encoder network that projects images into a smaller-sized latent space that is easier to sample from, and a decoder network that projects latent representations back to image space. The idea behind LDMs is to run a DM inside the latent space of a pretrained VAE. This allows for a reduction of the DMs computational demand by effectively compressing images to a smaller size through use of the VAE, while still preserving relevant details and thereby assuring high sample quality for large image sizes. In the context of radio astronomy, the only use of LDMs is by \cite{Sortino+2024}, who implement a model to generate single-source radio galaxy images.\\
In addition to implementing an LDM, we further develop a novel approach that allows us to break down the process of sampling large sky survey maps into several individual steps, thereby in principle facilitating the generation of sky maps of arbitrary size. This approach works by subsequently sampling individual patches of the map, within a sliding window that is moved with a stride of half the image size. At every step, the overlapping parts of previously sampled patches are passed as an additional input to the network. The model thereby samples neighboring images in a consistent fashion, which allows for step-wise sampling of large maps. We call this technique \textbf{S}liding \textbf{Wi}ndow \textbf{It}erative (\textsc{SWIIT}) sampling.\\
This paper is structured in the following way. In Section \ref{sec:LDMs}, we provide the necessary background to understand the theory behind LDMs. In Section \ref{sec:SWIIT-Sampling}, we explain our proposed \textsc{SWIIT}-sampling technique. Section \ref{sec:Training_Data} describes the collection and preparation of training images, Section \ref{sec:model-architecture-and-training} lays out the model architectures and training details. The performance of the trained Model and the quality of sampled images are evaluated in Section \ref{sec:results} and results are discussed in Section \ref{sec:discussion}.

\section{Latent diffusion models} \label{sec:LDMs}

DMs are a class of generative models that work by training a neural network (NN) to remove artificial Gaussian noise from images, and then using that network to synthesize novel realistic images from pure Gaussian noise in several steps of gradual denoising. While excelling in the quality and diversity of generated image data in particular \citep[e.g.][]{Dhariwal+21}, model training and sampling requires large computational resources. To alleviate this demand, \cite{Rombach+2021} combined DMs with a different class of generative models called autoencoders (AEs), and named the resulting method LDMs. AEs work by simultaneously training two NNs: An encoder that projects data samples into a smaller-sized latent space, and a decoder that reconstructs the original sample from the latent projection. The training loss, apart from enforcing accurate image reconstruction, typically constraints the representations of the training images to follow a distribution in latent space that can easily be sampled from. This is done in different ways, giving rise to multiple variants of the model class, like the widely used VAE \citep{Kingma+2019} or the vector-quantized variational autoencoder (VQ-VAE, \cite{van_den_Oord+2017}), which we use in this work. Synthetic samples can then be generated by first sampling from this latent space distribution and subsequently using the decoder to project those samples back to the original data space.\\
The combination of these two image generation methods hence involves applying a DM to sample within the latent space of a VQ-VAE. In this framework, the VQ-VAE is responsible for reconstructing fine-grained image details during decoding, while the DM operates on a compressed latent representation that primarily captures high-level semantic structure. The VQ-VAE therefore enables substantial image compression, which reduces the computational burden of the diffusion model and facilitates efficient high-quality image generation. In this section, we introduce the theoretical background necessary to understand the implementation of the LDM used in this work.

\subsection{Diffusion models}
DMs make use of a NN that is trained to remove Gaussian noise $\mathbf{z}$ of varying magnitudes $\sigma$ that is artificially added to images $\mathbf{x}$ during training. In doing so, the network implicitly learns the reverse of a process by which an image is gradually diffused with Gaussian noise. The latter is typically called the forward process, and its reverse the backward process. The forward process is described by a Markov Chain of length $T$, where at every step $t$ a small amount of Gaussian noise is added, such that
\begin{equation}
    \mathbf{x}_t = \mathbf{x} + \sigma_t \cdot \mathbf{z}_t,
\end{equation}
where $\sigma_t$ is the noise level at step $t$ and $\mathbf{z}_t$ is sampled from the standard normal distribution $\mathcal{N}\left(0, \mathbb{I}\right)$.
The corresponding reverse process is then used for image generation by starting with a seed image of pure Gaussian noise $\mathbf{x}_\mathrm{T} \coloneqq \sigma_\mathrm{T} \cdot \mathbf{z}$ with $\mathbf{z} \sim \mathcal{N}\left(0, \mathbb{I}\right)$ and gradually removing small amounts of noise in $T$ steps, which results in a novel image that resembles a sample drawn from the training distribution. More specifically, given a denoiser NN $D_\theta(\mathbf{x}, \sigma)$ that is trained to predict the corresponding denoised image from a noisy image $\mathbf{x}$ at noise level $\sigma$, the reversed process at step $t$ can be approximated as
\begin{equation}
    \mathbf{x}_{t-1} \approx \mathbf{x}_{t} - (\sigma_t - \sigma_{t-1}) \cdot\frac{\mathbf{x}_t - D_\theta(\mathbf{x}_t, \sigma_t)}{\sigma_t}.
\end{equation}
This approximation can be further improved by using a second network evaluation at every step. The detailed sampling procedure is explained in \cite{VicanekMartinez24}, and is the same as used for the DM in this work. The values for $T$ and $\{\sigma_t \,|\, t \in [0, T]\}$ are sampling hyperparameters, our choices are listed in Appendix \ref{app:architecture_and_training:DM}.

\subsection{Guided diffusion}
The NN can be trained with additional inputs $c$ of any shape, such as class labels or parameters that characterize the information contained on the image. This use of additional input information is referred to as conditioning, and DMs achieve high quality in conditional sampling through a technique called classifier-free guidance \citep{Ho&Salimans22}. For this, the conditioning input is randomly dropped out during training and replaced by a null input $\varnothing$ that has no effect on the network. The denoiser is then evaluated as a linear combination $\widetilde{D}_\theta$ of the conditioned and unconditioned network like
\begin{equation}
    \widetilde{D}_\theta(\mathbf{x}, \sigma_t \,|\, c) = (1 + \omega) \cdot D_\theta(\mathbf{x}, \sigma_t \,|\, c) + \omega \cdot D_\theta(\mathbf{x}, \sigma_t \,|\, \varnothing), 
\end{equation}
where the guidance strength $\omega$ is a sampling hyperparameter.

\subsection{Vector-quantized variational autoencoder}\label{sec:LDMs:VQVAE}
The following description of the VQ-VAE is based on the description provided in \cite{Rombach+2021}, which builds on the work of \cite{van_den_Oord+2017}. An autoencoder consists of two networks, the encoder $\mathcal{E}$ and the decoder $\mathcal{D}$. The encoder projects an input image $\mathbf{x}$ onto a smaller-sized latent representation $\mathbf{r}$, whereby the image is downsampled by a factor $f$, typically a power of two with $f = 2^m,\; m \in \mathbb{N}$. The decoder produces a reconstruction $\tilde{\mathbf{x}}$ of the input image from the latent, such that
\begin{equation}
    \tilde{\mathbf{x}} = \mathcal{D}(\mathbf{r}) = \mathcal{D}(\mathcal{E}(\mathbf{x})). \label{eq:autoencoder}
\end{equation}
This allows for compression of the image while still retaining all necessary information to fully reconstruct the original input.\\
In order to regularize the latent space, the possible pixel values that $\mathbf{r}$ can adopt are fixed to a discrete set of code vectors $\mathcal{C} = \{c_\mathrm{i}\}$, which is called the codebook. The code vectors $c_\mathrm{i}$ thereby correspond to the possible components, i.e. pixel values $r_\mathrm{i}$ of $\mathbf{r}$ of the latent images, and hence have the dimensionality of the number of channels in the latent space. This form of latent space regularization is called vector quantization, and it is carried out by mapping the pixels of the encoder output onto their nearest neighbor in the codebook. The quantized latent $\mathbf{q}(\mathbf{r})$ is hence obtained as
\begin{equation}
    q_\mathrm{i} = \mathop{\mathrm{argmin}}_{c_\mathrm{j} \,\in\, \mathcal{C}} \, \lVert c_\mathrm{j} - r_\mathrm{i} \rVert_2,
\end{equation}
where $\lVert \cdot \rVert_2$ is the $L_2$ norm. Eq. \eqref{eq:autoencoder} then becomes
\begin{equation}
    \tilde{\mathbf{x}} = \mathcal{D}(\mathbf{q}(\mathbf{r})) = \mathcal{D}(\mathbf{q}(\mathcal{E}(\mathbf{x}))). \label{eq:vq-autoencoder}
\end{equation}

The size $\left| \mathcal{C} \right|$ of the codebook is a fixed hyperparameter and the values of $c_\mathrm{i}$ are initialized at random and learned during training.

\subsection{Latent diffusion}
A latent diffusion model combines both approaches by first training an autoencoder on a set of images, and then training a DM on the latent representations of those images as obtained from the trained encoder. Synthetic samples are then generated by first using the DM to sample latent representations, and then using the decoder to reconstruct corresponding image samples. In this work, first a VQ-VAE is trained on the image training dataset. Subsequently, the trained encoder network is employed to create another training set containing the corresponding pre-quantized latent representations of the images, which are then used to train the DM denoiser network. In this context, the quantization operation is interpreted as the first layer of the decoder. Therefore, the DM denoiser is trained on the pre-quantized latents and the DM-generated latent samples are first quantized before being passed to the decoder.

\section{Sliding window iterative sampling} \label{sec:SWIIT-Sampling}
LDMs produce images of a fixed size that has an upper limit imposed by model complexity and computing demand, as laid out in Section \ref{sec:Introduction}. To facilitate sampling of sky maps at arbitrary sizes, we develop an approach that we call \textsc{SWIIT}-sampling. This approach is not limited to LDMs, and can be implemented for a pure DM as well.\\

\subsection{Method and implementation}
\begin{figure*}[h!]
    \centering
    \begin{subfigure}[h]{17.5cm}
        \centering
        \includegraphics[width=\hsize]{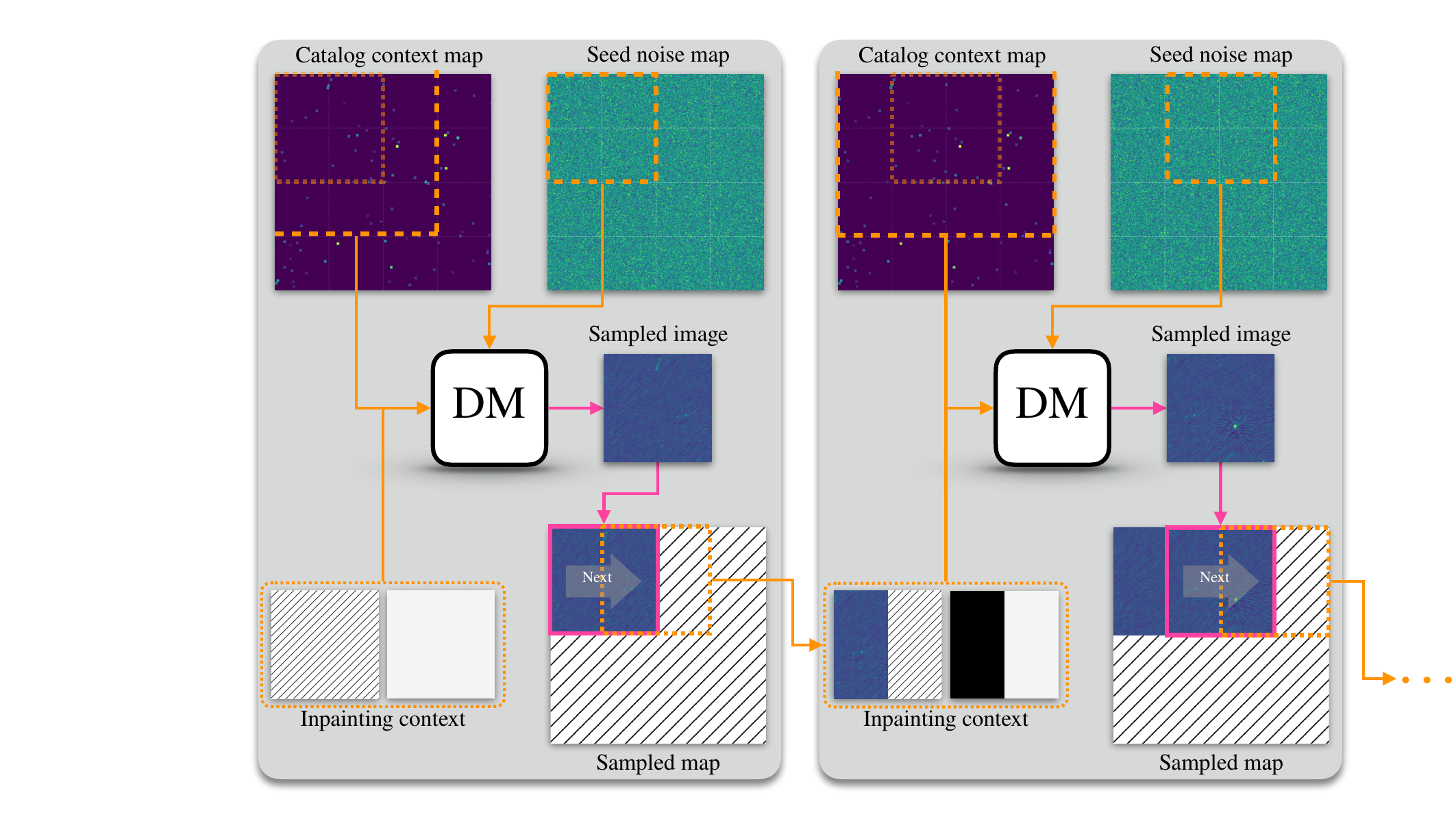}
        \caption{First steps of \textsc{SWIIT}-sampling.}
        \label{fig:sub:SWIIT_sampling_A}
    \end{subfigure}
    \begin{subfigure}[h]{17.5cm}
        \centering
        \includegraphics[width=\hsize]{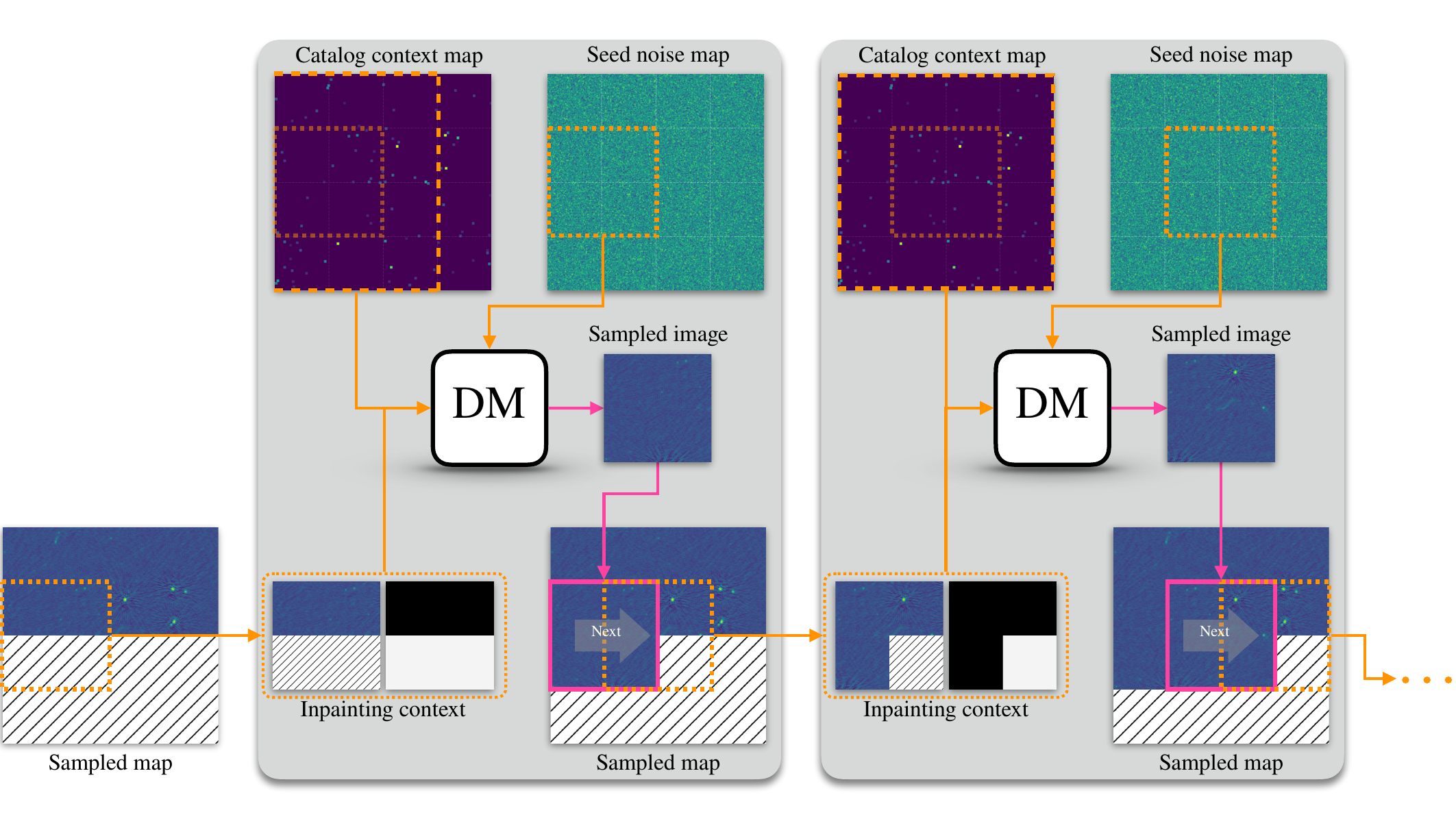}
        \caption{Intermediate steps of \textsc{SWIIT}-sampling.}
        \label{fig:sub:SWIIT_sampling_B}
    \end{subfigure}
    \caption{Schematic representation of the \textsc{SWIIT}-sampling procedure with a final map size of \SI{768}{\px} side length, corresponding to a total of 9 sampling steps. Hatch lines indicate empty regions. The inpainting masks are represented in black for values of 0 and white for values of 1. Note that, although this procedure is done with latent images, we show image data for illustration purposes.}
    \label{fig:SWIIT_sampling}
\end{figure*}

\begin{figure}[h!]
    \begin{subfigure}[h]{\hsize}
        \centering
        \resizebox{\hsize}{!}{
        \includegraphics[width=\hsize]{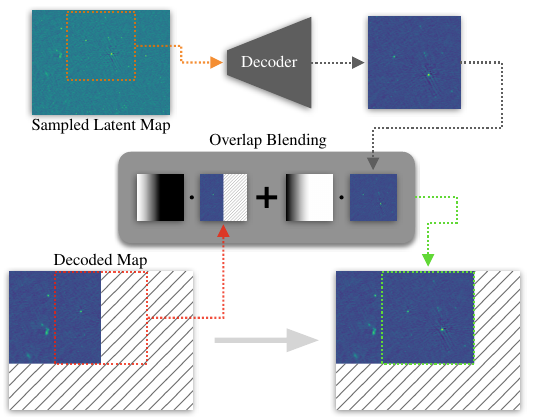}}
        \subcaption{Blended decoding procedure, showing blend between first and second step of decoding.}
        \label{fig:sub:decode_blending_1}
    \end{subfigure}
    \begin{subfigure}{\hsize}
        \centering
        \includegraphics[width=\hsize]{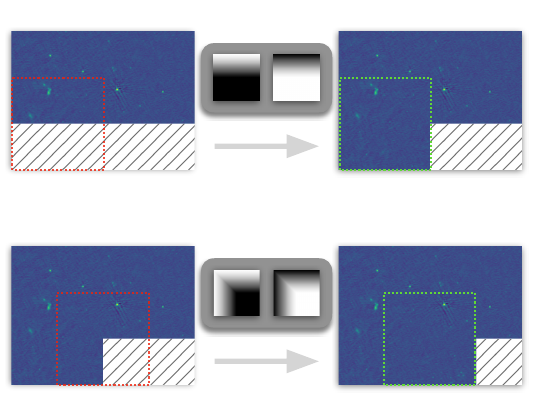}
        \subcaption{Other steps of decoding shown with the respective blending gradients.}
        \label{fig:sub:decode_blending_2}
    \end{subfigure}
    \caption{Illustration of the blended decoding procedure, showing examples of different stages with the respective blending gradients as grayscale images, where black represents 0 and white represents 1.}
    \label{fig:decode_blending}
\end{figure}

For our approach, we train the diffusion model to not only sample images entirely, but to also reconstruct missing pieces of images that are partly visible to the network. This kind of image generation is generally known as image inpainting, a task that LDMs have been shown to be suitable for \citep{Rombach+2021}. This capacity of the model allows us to sequentially sample radio maps of arbitrary sizes by using a sliding window, where we first sample an image entirely, then move the sliding window and sample another overlapping image, with the overlapping parts of the previous sample visible to the network, and so on.\\
In order to facilitate a high level of control over the DM’s sampling behavior, we train the denoiser with an additional conditioning input that provides information regarding positions and properties of the sources present on any training image. This input takes the form of an image-shaped tensor that encodes source positions, as well as their brightness and size, through pixel values at the corresponding locations. These values are derived from a source catalog, see Section \ref{sec:Training_Data} for details. We therefore refer to this representation as the catalog context vector. The context vector has twice the spatial extent of the image, allowing sources in the surrounding region to be incorporated into the conditioning. In addition, we provide an image-sized binary inpainting mask that indicates the region to be generated - i.e., denoised during training and inpainted during sampling - along with the complementary, fixed portion of the image. We refer to this latter pair of inputs as the inpainting context.\\
The procedure of \textsc{SWIIT}-sampling is illustrated in Figure \ref{fig:SWIIT_sampling}. A synthetic radio sky map is generated from a catalog context map of predefined size, holding the information of sources that are to be present on the sampled map. We first sample Gaussian seed noise with the shape of the corresponding latent map, i.e. the downsampled size of the map to be sampled. We then start at the top-left corner and sample the first latent image entirely with no inpainting, using both the seed noise and the corresponding source context vector at that position. Next, we move the sliding window to the right by half a latent image width to produce the next sample. For this, we use the right half of the first sampled latent as the left half of the inpainting context, and in turn generate the right half of the latent at the current sliding window position. This is illustrated in Figure \ref{fig:sub:SWIIT_sampling_A}. The same process is continued until the desired map width is reached. The sliding window is then moved back all the way to the left, and down by half an image width, see Figure \ref{fig:sub:SWIIT_sampling_B}. The procedure is repeated until reaching the intended size. This means that for the top left corner, the entire latent is sampled with no inpainting, for the rest of the top row, only the right half is sampled at each step, for the left column, the bottom half is sampled at every step, and for all other steps of the sampling procedure only the bottom right quadrant is sampled. This procedure results in a latent map of desired size. As a final step, this  latent map is decoded into image space using the VQ-VAE decoder. This is again done in a sliding-window convolutional fashion, using a stride of half the latent image size. Overlapping portions of subsequent decoding steps are blended together with a smooth gradient, as shown in Figure \ref{fig:decode_blending}, that is shaped according to the overlapping area between the images. This is implemented as an array that smoothly interpolates between values of 0 and 1, which is used to scale the corresponding latent patch pixel-wise. This prevents edge-like decoding artifacts from appearing on the final map. The result of this procedure is a sampled radio map of desired size, with control over source positions and properties. \\
To train the network for optional image inpainting, we pass partly masked input images together with the mask as inpainting context to the network during training. These inpainting masks cover different combinations of image quadrants, corresponding to the four different cases that arise throughout the sampling procedure explained above. The four possible masks are illustrated by the different inpainting context shapes in Figures \ref{fig:sub:SWIIT_sampling_A} and \ref{fig:sub:SWIIT_sampling_B}. On the inpainting mask, a value of 1 indicates that the part of the image should be reconstructed by the network, meaning that at inference time this part of the image will be sampled. Consequently, the input image is multiplied with the inverse of that mask before being passed to the network as inpainting context. During sampling, at each denoising step $t$ the part of the image not covered by the sampling mask is replaced with the noisy original input, i.e.
\begin{equation}
    \mathbf{x}_{t} \leftarrow 
    \begin{cases}
        \mathbf{x}_{t} & \text{Sampled parts},\\
        \mathbf{x}_\mathrm{Inp} + \sigma_t \cdot \mathbf{z} & \text{Non-sampled parts},
    \end{cases}
\end{equation}
where $\mathbf{x}_\mathrm{Inp}$ denotes the image of the inpainting context.\\
As is also illustrated in Figure \ref{fig:SWIIT_sampling}, since for any sampled image the catalog context vector covers neighboring regions as well, part of this information is not available for sampling steps at the edge of the sampled map. In those cases, the unavailable parts of the catalog context vector are blanked, i.e. set to zero. These cases are accounted for during training by randomly blanking corresponding parts of the catalog context vector, more details are given in Section \ref{sec:training:DM} and in Appendix \ref{app:architecture_and_training:DM}.

\subsection{Inference}
For inference, the only required input is the target size of the sampled map, specified by its height and width in pixels, where one pixel corresponds to \ang{;;1.5}. Providing a catalog context is optional, since the model is also trained for context-free generation through context dropout; however, passing context facilitates control over the generated samples. The context is constructed from an input catalog covering the area of the desired map size and containing source positions, total flux in \si{\milli Jy}, peak flux in \si{\milli Jy \per beam}, and major axis, or an analogous size parameterization, in \si{arcsec}. Such a catalog may be obtained in several ways. Possible sources include catalogs derived from real observed radio maps or catalogs simulated with frameworks such as \textsc{TRECS} \citep{Bonaldi+2019}. Alternatively, three-dimensional histograms of the relevant source properties may be constructed from existing survey catalogs and sampled to generate a desired number of sources. Source positions can then be assigned using a custom prescription, for example a uniform random distribution.\\
The catalog must subsequently be converted into a context vector as described in detail in Section \ref{sec:Training_Data}, yielding a single input context vector that corresponds to the full desired map size. If reproducibility is required, a seed noise map for the diffusion process may also be provided manually. This noise map should consist of standard normal Gaussian noise and have a spatial resolution that is a factor of four smaller than the target map size, corresponding to the downsampling factor between image space and latent space.

\begin{figure}
    \centering
    \includegraphics[width=\hsize]{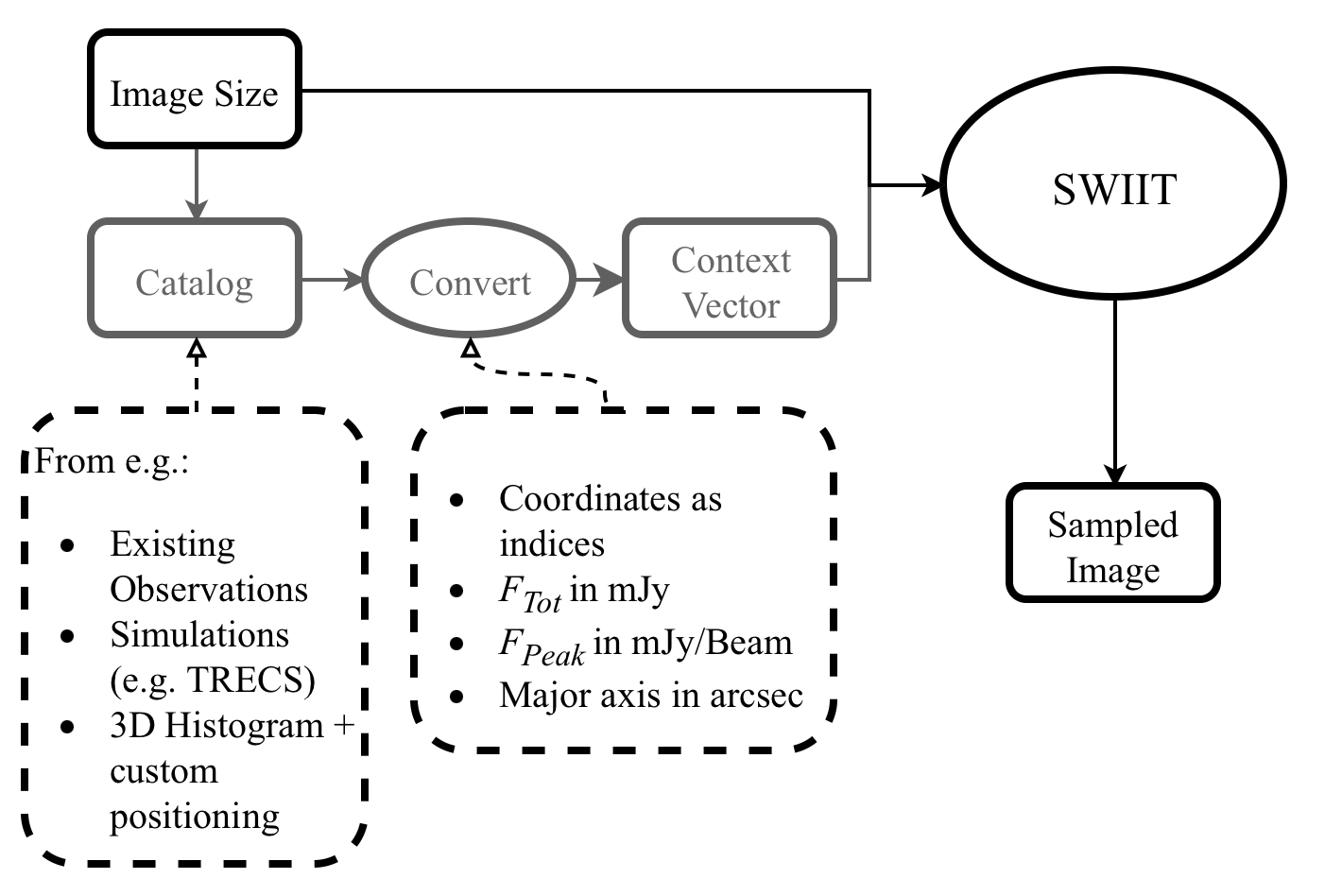}
    \caption{Schematic diagram describing the necessary inputs for \textsc{SWIIT}-Sampling. Rectangular boxes represent data objects, elliptical boxes represent processing steps, dashed lines indicate additional information. Grey elements are optional.}
    \label{fig:sampling_schematic_diagram}
\end{figure}

\section{Training data} \label{sec:Training_Data}
\begin{figure}[h!]
    \centering
    \begin{subfigure}[t]{0.49\hsize}
        \includegraphics[width=\hsize]{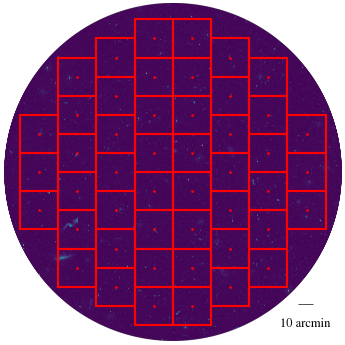}
        \subcaption{Tiling pattern on the mosaic.\\}
        \label{fig:cutout_tiling:tiles}
    \end{subfigure}
    \begin{subfigure}[t]{0.49\hsize}
        \includegraphics[width=\hsize]{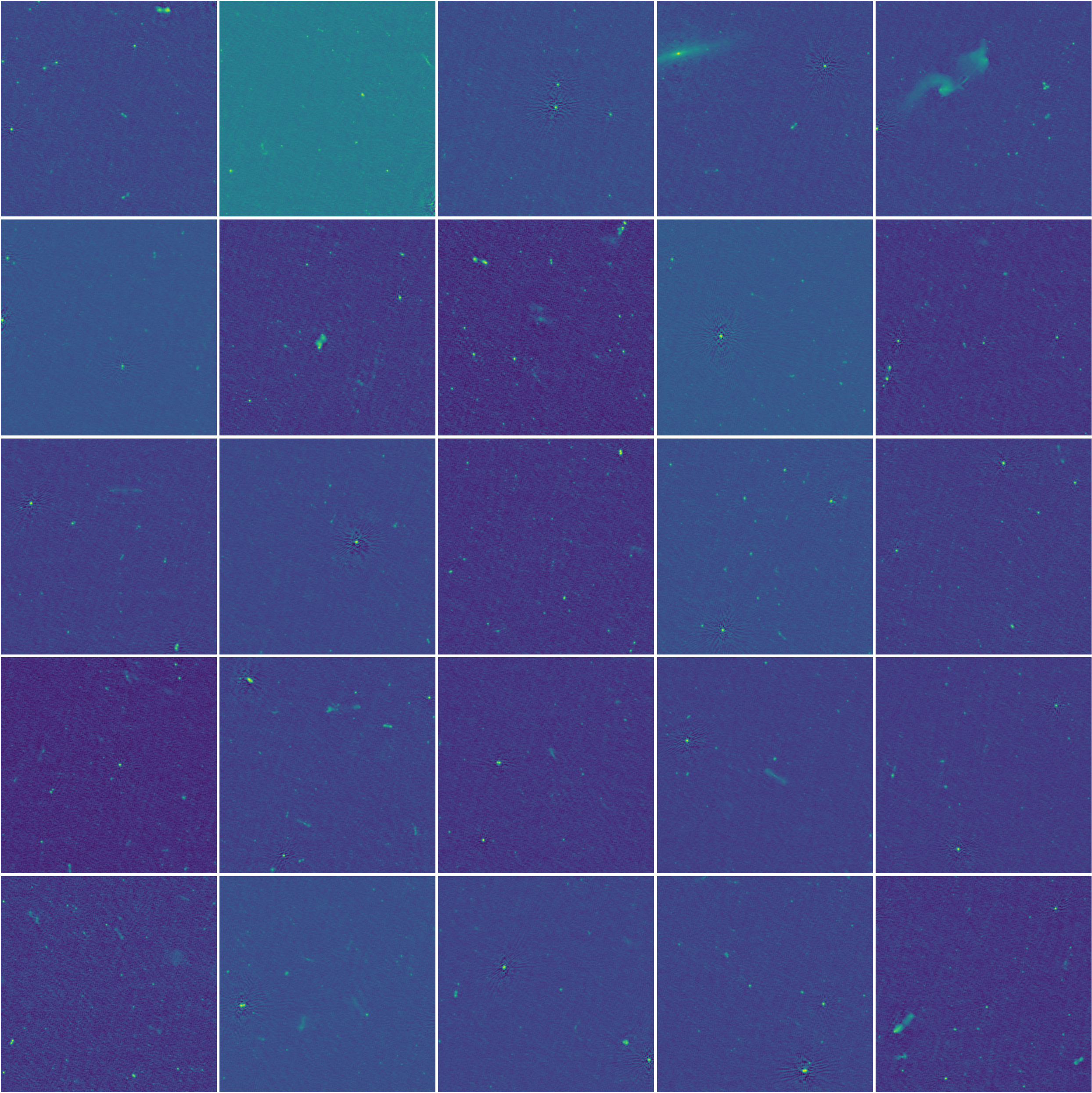}
        \subcaption{Selection of resulting cutouts with \SI{1024}{px} (\ang{;25.6;}) side length.}
        \label{fig:cutout_tiling:cutouts}
    \end{subfigure}
    \caption{Illustration of the cutout extraction for a single LoTSS-DR3 mosaic, showing the tiling pattern (left) and the resulting cutouts after pixel scaling (right).}
    \label{fig:cutout_tiling}
\end{figure}

\begin{figure}
    \centering
    \includegraphics[width=\hsize]{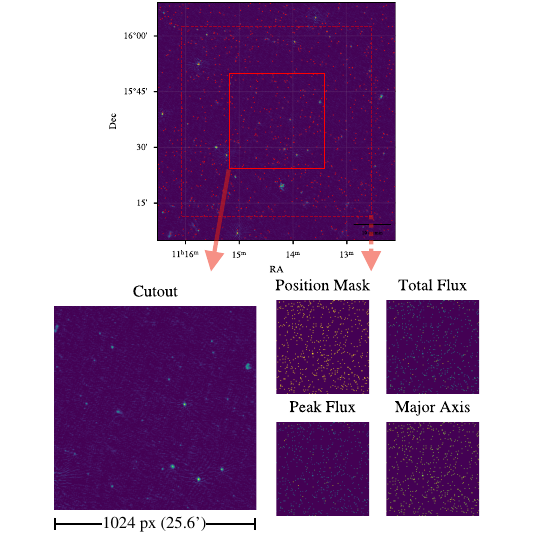}
    \caption{Illustration showing the extraction of training data. The top plot shows part of a LoTSS-DR3 mosaic, with source positions scattered as red dots. The inner solid-line square indicates the extracted image cutout. The outer dashed-line square indicates the boundaries for extracting the catalog context, for which the four channels are shown on the bottom right.}
    \label{fig:cutout_extraction}
\end{figure}

The training data for our LDM is obtained from observations made with the Low Frequency Array (LOFAR) radio telescope \citep{van_Haarlem2013}. Individual cutouts are extracted from publicly available mosaic images issued with the third data release of the LOFAR Two-Metre Sky Survey \citep[LoTSS DR3,][]{Shimwell+2026}. These maps cover 88\% of the northern sky in a band of 120 to \SI{168}{\mega \hertz} with a central frequency of \SI{144}{\mega \hertz}. The mosaics have a resolution of \ang{;;6} above declination \ang{10} and \ang{;;9} resolution below. The data release also features a source catalog, obtained with the Python Blob Detector and Source Finder \citep[\textsc{PyBDSF},][]{pyBDSF}, which we use to create our catalog context arrays. \textsc{PyBDSF} is a software that detects radio sources and models their morphologies as composites of two-dimensional Gaussians. Among many other quantities, this catalog lists for every source the integrated and peak flux densities, as well as the size parametrized as the major-axis calculated through image moment analysis on the source model.\\
Training images are obtained by dividing individual LoTSS mosaics into square-shaped tiles as shown in Figure \ref{fig:cutout_tiling} and saving them as arrays. The procedure for determining the optimal tiling pattern is described in Appendix \ref{app:mosaic_tiling}. We create a dataset with cutouts of size \SI{1024}{\px} (\ang{;25.6;}). While the LDM is trained on images of \SI{512}{\px} (\ang{;12.8;}), the larger-sized cutouts facilitate random augmentations that are described in Section \ref{sec:training:DM}. For some mosaics, parts of the image hold invalid pixel values. For contiguous regions of no more than 8 invalid pixels, those pixels are iteratively filled with the mean value of their valid 8-connectivity neighboring pixels. Cutouts containing regions of more than 8 invalid pixels are entirely discarded. Pixel values of extracted cutouts are scaled with a logarithmic function $\gamma(x)$ described in Appendix \ref{app:pixel_scaling}. This scaling is applied to transform the asymmetric distribution of pixel values into a distribution close to zero-mean and unit variance, which is better suited for machine learning applications.\\
In addition to the training images, we extract information on the positions, fluxes and sizes of the sources present on a cutout and its immediate surroundings outside of the cutout field of view (FOV). This information is then encoded as a catalog context vector, thereby put into a format that can be passed as a conditioning input to the denoiser NN of the DM. This encoding is illustrated in Figure \ref{fig:cutout_extraction}. First, for any individual cutout, we filter the DR3 source catalog for sources that fall within the boundaries of the cutout and its surroundings. This boundary is defined as a square around the cutout center with double the size of the cutout. During training, this allows for inclusion of sources that lie outside the FOV, but whose presence still potentially influences the image through spillover artifacts. This is required for consistency between adjacent samples of the same sampled map. For all sources within the boundaries, we extract the source positions, as well as the integrated flux, peak flux, and major-axis values from the catalog. The values remain unscaled and take units of \si{\milli Jy}, \si{\milli Jy \per beam} and \si{arcsec}. Using these parameters, we create a 4-channel image of double the cutout size. The first channel is set to 1 at pixels corresponding to source positions and -1 otherwise. The second to fourth channel are filled with the extracted values of the three respective parameters, again at the pixel corresponding to the source position, and set to 0 otherwise. We call this array the catalog context array. This array will be used as contextual input to the DM's denoiser network, which operates on the latent representations that are reduced in size by $f=4$. Hence, as a final step, the catalog context arrays for every cutout are downsampled by a factor of 4 via sum-pooling, whereby empty pixels in the position channel are temporarily set to 0 instead of -1.\\
The described procedure results in \num{152098} cutouts of $\SI{1024}{px}$. When selecting for those with $\ang{;;6}$ resolution, this number reduces to \num{113184}. We separate the data set into training, validation and test splits at proportions of \num{0.8}, \num{0.1} and \num{0.1}, randomly picked across all DR3 mosaics.

\section{Model architecture and training} \label{sec:model-architecture-and-training}
\begin{figure}
    \centering
    \includegraphics[width=\hsize]{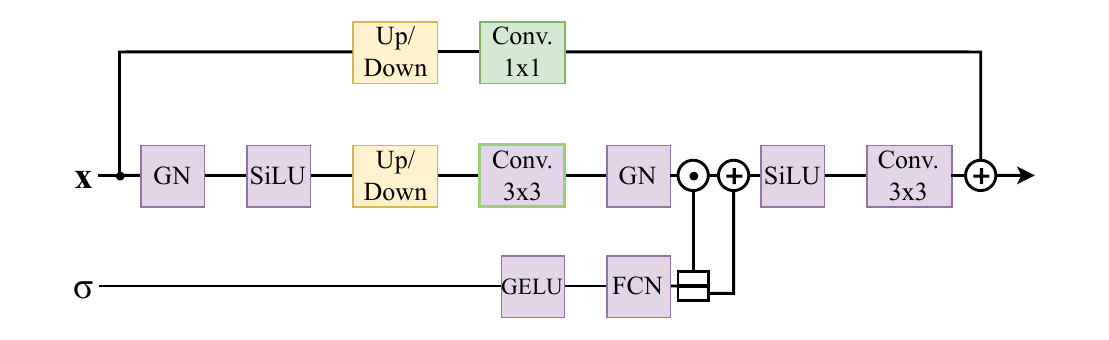}
    \caption{Architecture of the CNN block, including group normalization (GN) and sigmoid linear units (SiLU). Conditional inputs are passed through a Gaussian error linear unit (GELU) and a fully-connected network (FCN). Optional elements used in different parts of the networks are colored in agreement with Figures \ref{fig:VQ-VAE_architecture} and \ref{fig:UNet_architecture}. The first 3$\times$3-convolution is always present, only the included change in channel dimensions is optional. The noise level injection is only present in the U-Net, not in the VQ-VAE. This figure is adopted from \cite{VicanekMartinez24}.}
    \label{fig:cnn_block}
\end{figure}

The encoder-decoder architecture used for the VQ-VAE is nearly identical to the U-Net \citep{Ronneberger+2015} employed for the denoiser network of the DM, with only a few punctual differences. We use the same architecture employed previously in \cite{VicanekMartinez24}, where it is described in great detail. In this section, we give a brief summary and emphasize the differences and changes or new additions. A more detailed technical description with full listing of hyperparameter choices is given in Appendix \ref{app:architecture_and_training}.\\
Both networks are composed of individual blocks that can be slightly altered depending on their position in the network. Every block features two convolutional neural network (CNN) layers with residual connections, and some feature optional self-attention layers, see Figure \ref{fig:cnn_block}. The blocks can, also optionally, contain a down- or upsampling module that halves or doubles the number of pixels in the feature map. This gives rise to different resolution levels throughout the network, and each resolution level holds two CNN blocks and optional down- or upsampling blocks. Finally, the block can also optionally increase or reduce the number of channels in the output feature map.

\subsection{Vector-quantized variational autoencoder}
\begin{figure}
    \centering
    \includegraphics[width=\hsize]{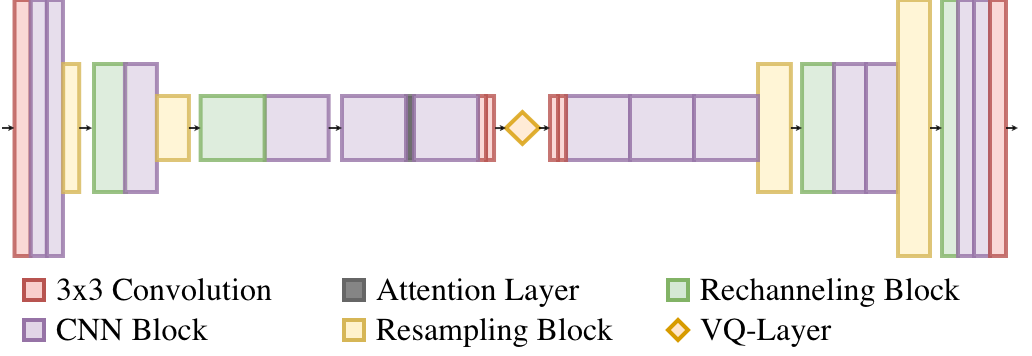}
    \caption{Architecture of the VQ-VAE Encoder-Decoder network. Block heights indicate resolution, widths indicate channel depth.}
    \label{fig:VQ-VAE_architecture}
\end{figure}

The VQ-VAE architecture is illustrated in Figure \ref{fig:VQ-VAE_architecture}. Both encoder and decoder have three resolution levels with no self-attention layers. The encoder at its highest resolution level and the decoder at its lowest level are preceded by an initial CNN layer. At the lowest resolution layer, the encoder has an additional bottleneck sequence of two CNN blocks with a self-attention layer in between. After those, the mapping to pre-quantized latent space happens through two final CNN layers. The decoder follows the exact same architecture in reverse order, with upsampling instead of downsampling layers \footnote{For practical reasons, our decoder has three CNN blocks at every resolution level, rather than two like the encoder. This is a consequence of adapting the implementation of the U-Net architecture for the VQ-VAE.}. Vector quantization happens in-between.\\
For training the VQ-VAE, we adopt several strategies from \cite{Huh+2023}, which in turn builds on the work presented in \cite{van_den_Oord+2017}. The network is optimized with several different loss terms. The first is the reconstruction loss, which compares the input image to the reconstructed output image as 
\begin{equation}
    \mathcal{L}_\mathrm{Rec} = \lVert \tilde{\mathbf{x}} - \mathbf{x} \rVert_1, \label{eq:VQ-VAE_RecLoss}
\end{equation}
where $\lVert \cdot \rVert_1$ is the $L_1$ norm and all other definitions are consistent with Section \ref{sec:LDMs:VQVAE}. In order to optimize for high reconstruction accuracy also at large pixel values, we choose to include another term for the reconstruction loss using unscaled pixels. With the inverse pixel scaling function $\gamma^{-1}$ defined in Eq. \eqref{eq:pixel_scaling_inverse}, this term reads
\begin{equation}
    \mathcal{L}_\mathrm{Rec, unsc} = \lVert \gamma^{-1}(\tilde{\mathbf{x}}) - \gamma^{-1}(\mathbf{x}) \rVert_1. \label{eq:VQ-VAE_RecLossUnsc}
\end{equation}
In addition to reconstruction loss, we train a discriminator NN in parallel to the VQ-VAE. The discriminator is trained to distinguish between real training images and those reconstructed by the decoder, thereby introducing an additional optimization objective for the VQ-VAE to produce realistic images in a fine-grained way that is hard to optimize for by exclusively using the reconstruction loss. The discriminator network is a simple five-layer CNN that takes an image as input and and outputs a single scalar intended to label original training images as \num{+1} and reconstructed images as \num{-1}. For a batch $\mathcal{B} = \mathcal{B}_\mathrm{Real} \cup \mathcal{B}_\mathrm{Rec}$ containing real and reconstructed images, the discriminator $\mathcal{C}$ is optimized with the loss function
\begin{equation}
    \mathcal{L}_\mathrm{Disc} = 
    \begin{cases}
        \relu\left(1 - \mathcal{C}(\mathbf{x}) \right) & \mathbf{x} \in \mathcal{B}_\mathrm{Real},\\
        \relu\left(1 + \mathcal{C}(\mathbf{x}) \right) & \mathbf{x} \in \mathcal{B}_\mathrm{Rec},\\
    \end{cases}
\end{equation}
where $\relu(x) \coloneqq \max(0, x)$. In turn, the VQ-VAE training objective is complemented with the generator loss term
\begin{equation}
    \mathcal{L}_\mathrm{Gen} = - \mathcal{C}(\tilde{\mathbf{x}}),
\end{equation}
which penalizes reconstructions correctly identified as such by the discriminator.
Finally, the quantization layer is trained to learn an optimized set of code vectors through the commitment loss $\mathcal{L}_\mathrm{Commit}$, which incentivizes alignment between the encoder output $\mathbf{r}$ and the quantized latents $\mathbf{q(\mathbf{r})}$, reading
\begin{equation}
    \mathcal{L}_\mathrm{Commit} = (1 - \beta) \cdot (\mathbf{r} - \mathrm{sg}[\mathbf{q(\mathbf{r})}])^2 + \beta \cdot (\mathrm{sg}[\mathbf{r}] - \mathbf{q(\mathbf{r})})^2,
\end{equation}
where $\beta$ is a hyperparameter and $\mathrm{sg}[\cdot]$ is the stop-gradient operator. The latter is defined as identity at forward computation but with zero partial derivatives, hence providing the effect that the operand receives no gradient update. Therefore, the first term in $\mathcal{L}_\mathrm{Commit}$ updates the encoder network to output latent pixel values closer to the code vectors, whereas the second term updates the codebook towards more favorable code vectors.\\
In total, the VQ-VAE is optimized with a linear combination
\begin{equation}
    \begin{split}
        \mathcal{L}_\mathrm{VQ-VAE} &= \alpha_\mathrm{Rec} \mathcal{L}_\mathrm{Rec} + \alpha_\mathrm{Rec,unsc} \mathcal{L}_\mathrm{Rec,unsc}\\ &+ \alpha_\mathrm{Gen} \mathcal{L}_\mathrm{Gen} + \alpha_\mathrm{Commit} \mathcal{L}_\mathrm{Commit},
    \end{split}
\end{equation}
where the different $\alpha_\mathrm{i}$ are training hyperparameters.\\
The use of three resolution levels, where the size of the feature map is halved between each, corresponds to $m = 2$ with a downsampling of $f=2^m = 4$. We use a latent dimension of 3 channels. Although the LDM will eventually operate on images of \SI{512}{\px}, the architecture is agnostic to image size and performs equally well on images larger than the training examples. Hence, we train the model on \SI{256}{\px}-examples to reduce computational demands. In order to increase the effective number of training examples, we extract random crops from the \SI{1024}{\px}-cutouts during training time. We use only the training images with $\ang{;;6}$ resolution.

\subsection{Diffusion model}\label{sec:training:DM}
\begin{figure}
    \centering
    \includegraphics[width=\hsize]{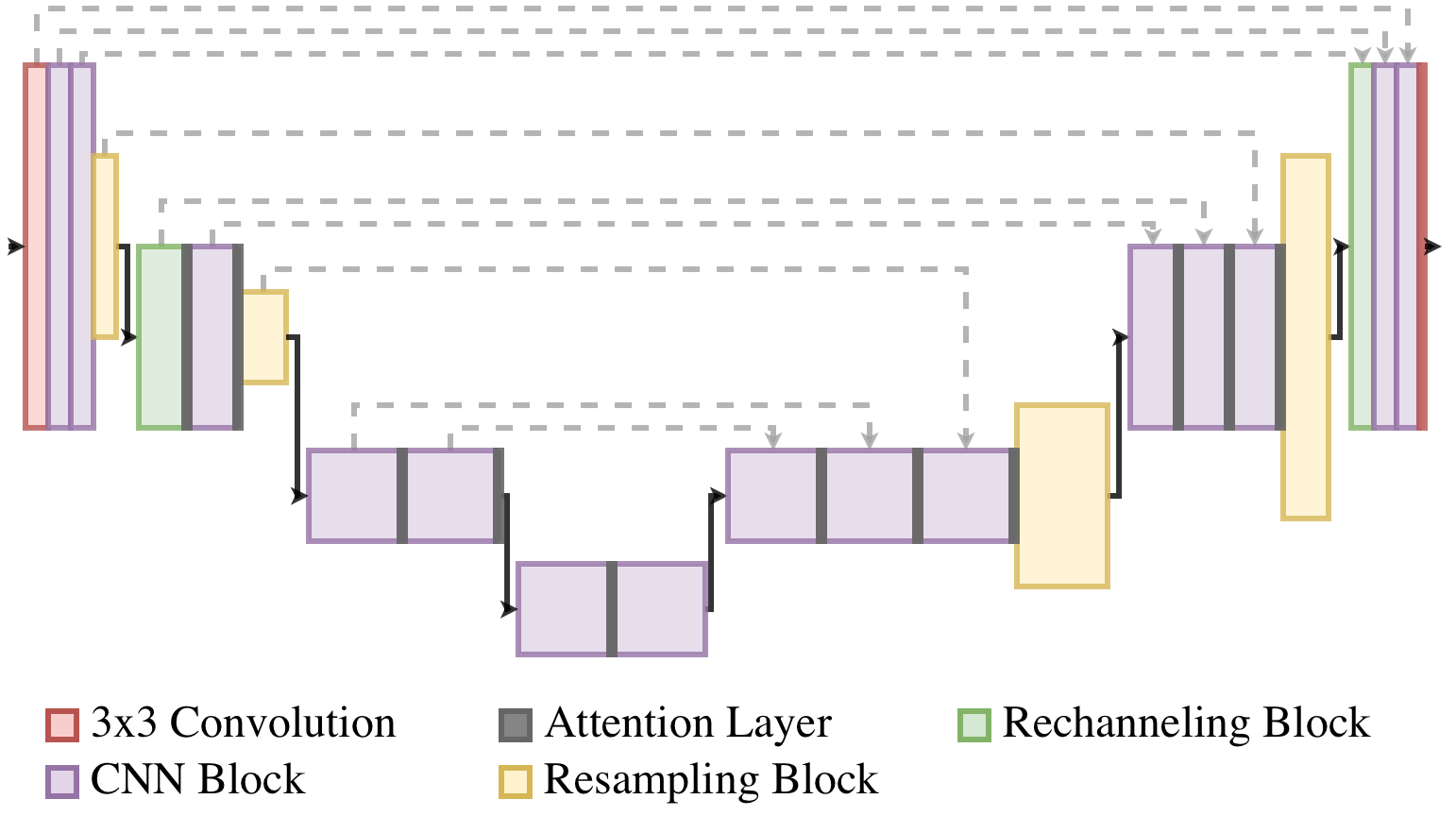}
    \caption{Architecture of the denoiser U-Net. Skip connections are indicated with dashed lines. Block heights indicate resolution, widths indicate channel depth.}
    \label{fig:UNet_architecture}
\end{figure}

The DM denoiser network is implemented as a U-Net, which largely follows the same architecture as the VQ-VAE encoder-decoder network, albeit with a few crucial differences. The architecture is illustrated in Figure \ref{fig:UNet_architecture}. The main difference is that the U-Net features additional residual connections between the encoder and the decoder part at levels of equal resolution, making it one consistent network rather than two separate components. Also, at the lowest resolution level the U-Net has a bottleneck layer of two CNN blocks with a self-attention layer in-between, and does not feature any additional CNN layers. The two levels of lowest resolution in the U-Net have self-attention layers between the CNN blocks. \\
Alongside the input image, the U-Net receives conditional inputs of different kinds. The diffusion noise level $\sigma$ is passed as a scalar value to a fully-connected network (FCN) before being injected into each of the CNN blocks between the first and second CNN layer through feature-wise linear modulation \citep[FiLM,][]{Perez+2018}, as shown in Figure \ref{fig:cnn_block}. The inpainting context is concatenated to the input image along the channel dimension. To facilitate learning of the complex relationship between the cutout and the extended catalog context array, which features sources outside the cutout FOV, we first pass the catalog context to a smaller U-Net that functions as an encoder for the catalog context vector and outputs a three-channel image of the same size, i.e. double the size of the latent image. This context encoding is then center-cropped to the size of the latent image and finally also concatenated to the input along the channel dimension.\\
The denoiser $D_\theta$ is trained with $L_2$ reconstruction loss. For an input image $\mathbf{x}$ and added noise $\mathbf{z} \sim \mathcal{N}(0, \sigma^2 \mathbb{I})$, the loss reads
\begin{equation}
    \mathcal{L} = c_\mathrm{out}(\sigma)^{-2} \cdot \lVert D_\theta(\mathbf{x} + \mathbf{z}, \sigma) - \mathbf{x}) \rVert^2_2, \label{eq:denoiser_loss}
\end{equation}
where $c_\mathrm{out}$ is a hyperparameter discussed in Appendix \ref{app:architecture_and_training:DM} that balances the loss magnitude over different values of $\sigma$. For any training step, the loss is calculated over the entire image, regardless of the inpainting context. We found this to be crucial for \textsc{SWIIT}-sampling.\\
Before training the denoiser, we first apply the trained VQ-VAE encoder on our image cutout dataset to create a corresponding latent image dataset. These latents have sizes of \SI{256}{\px}. We use only training images of \ang{;;6} resolution. To augment the effective number of images seen by the model during training, we again extract random crops of \SI{128}{px} at training time. This is done for the image cutouts and the catalog context array, where the crop is centered around the equivalent position but has double the size. The inpainting mask is randomly varied during training with uniform probabilities between four possible shapes. As laid out in Section \ref{sec:SWIIT-Sampling}, these correspond to sampling the entire image, sampling the right half, sampling the bottom half, and sampling the bottom right corner. To recreate the conditions of sampling the edges of a \textsc{SWIIT}-map, parts of the catalog context are randomly blanked in correspondence with the inpainting mask. Further details are given in Appendix \ref{app:architecture_and_training:DM}.

\section{Results} \label{sec:results}

\subsection{VQ-VAE for semantic image compression} \label{sec:results:vq-vae}
\begin{figure}[h!]
    \centering
    \includegraphics[width=\hsize]{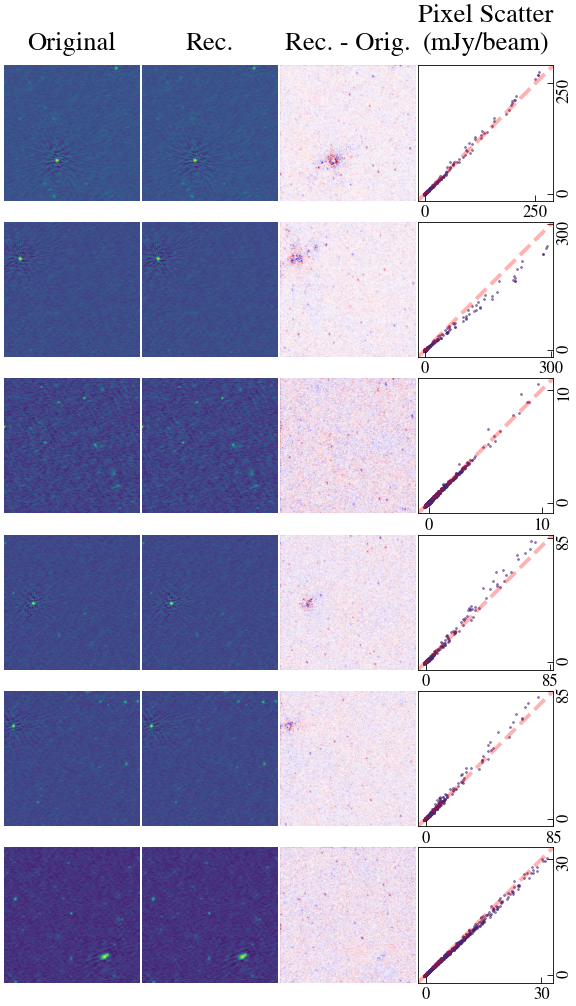}
    \caption{Examples of test cutouts in the first column and corresponding VQ-VAE reconstructions in the second column. Third column shows reconstruction difference, where the symmetric color scale is adjusted to 10\% of the maximum value of both the original and reconstructed image. Fourth column shows pixel-wise scatter plot of unscaled pixels, where axis limits are set for every image individually.}
    \label{fig:vae_examples}
\end{figure}

\begin{figure}[h!]
    \centering
    \includegraphics[width=\hsize]{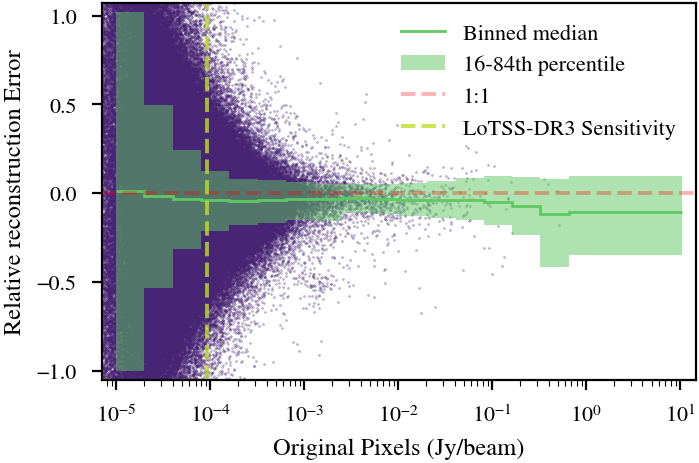}
    \caption{Relative per-pixel reconstruction error for all unscaled test cutouts combined. Bins are enforced to contain a minimum of \num{2500} data points. The vertical dashed line indicates the median sensitivity of \SI{92}{\micro\Jy\per\beam} reported in \cite{Shimwell+2026}. We scatter a random selection of \num{e6} data points for visualization.}
    \label{fig:vae_reconstruction_error}
\end{figure}

To assess the performance of the VQ-VAE, we use the entire test set  covering \num{11460} cutouts and generate the corresponding reconstructions. For each cutout, we use the central crop of \SI{512}{\px} side length.\\
A number of selected examples is shown in Figure \ref{fig:vae_examples}. Under visual inspection the reconstructions are practically indistinguishable from the original images. The third column in the figure shows the reconstruction difference, whereby the color range is scaled to the absolute maximum value of the input image to visualize the relative scale of deviations. The very few visible spots where reconstruction error is present are typically allocated in the vicinity of bright sources. The fourth column shows the per-pixel reconstruction scatter plot for unscaled pixels, i.e. input pixels before applying $\gamma$ and output pixels after applying $\gamma^{-1}$. Most pixels are strongly aligned with the line of equality, with visible deviations typically appearing for large pixel values. This is expected from the logarithmic shape of $\gamma$, where small reconstruction errors of scaled pixel values translate to increasingly larger errors in unscaled pixels at increasing pixel values.\\
To obtain a statistic of the the reconstruction error over the entire test set, we calculate the pair-wise relative reconstruction error $\delta$ between unscaled input pixels $x_\mathrm{In}$ and unscaled output pixels $x_\mathrm{Out}$ as 
\begin{equation}
    \delta = \frac{x_\mathrm{Out} - x_\mathrm{In}}{x_\mathrm{In}}.\label{eq:relative_reconstruction_error}
\end{equation}
The result is shown in Figure \ref{fig:vae_reconstruction_error}, with a random selection of scattered data points and statistics calculated over logarithmically spaced bins of input values. The median reconstruction error has a magnitude on the order of \num{e-2} for most of the bins, only reaching $\sim\num{e-1}$ at extremely high pixel values. Interestingly, there is a small bias towards smaller pixel values in the reconstructed images across the entire range. This can possibly be again related to the pixel scaling function $\gamma$, where positive reconstruction errors in unscaled images are reduced more strongly in magnitude by application of $\gamma^{-1}$ as compared to negative reconstruction errors in the same regime. A strong scatter of $\delta$ is present at very low pixel values around and below the median LoTSS-DR3 sensitivity \citep{Shimwell+2026}, where the signal is increasingly dominated by noise. This is expected for two reasons. First, for very small values of $x_\mathrm{In}$, the relative error $\delta$ in Equation \eqref{eq:relative_reconstruction_error} can blow up in magnitude even at small absolute reconstruction errors. Second, in the regime of high-level image detail with small pixel values and little semantic information, we expect the decoder to generate pixels optimized more strongly for the generator loss term, given the small absolute contribution to the reconstruction loss, i.e. to make reconstructions more "realistic" than "accurate". 

\subsection{Latent diffusion model} \label{sec:results:ldm}
\begin{figure}
    \centering
    \includegraphics[width=0.8\hsize]{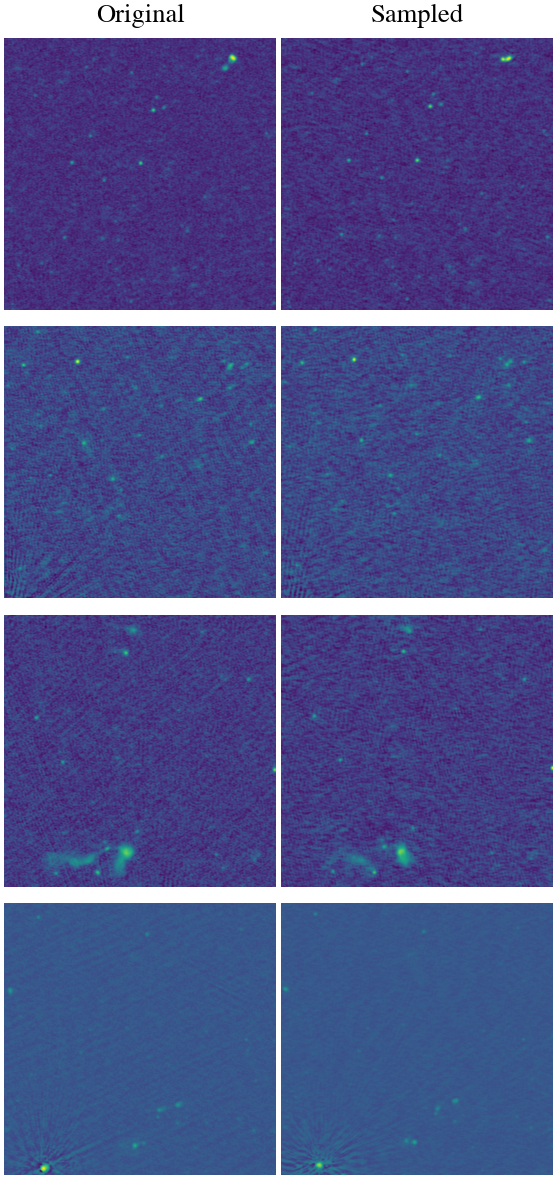}
    \caption{Example pairs of test cutouts with \SI{512}{px} (\ang{;12.8;}), showing the original cutout on the left and a generated sample on the right. Note that we do not expect an exact reconstruction, but a realistic image with sources at the same locations, of same brightness and size.}
    \label{fig:ldm_examples}
\end{figure}

\begin{figure*}
    \centering
    \includegraphics[width=180mm]{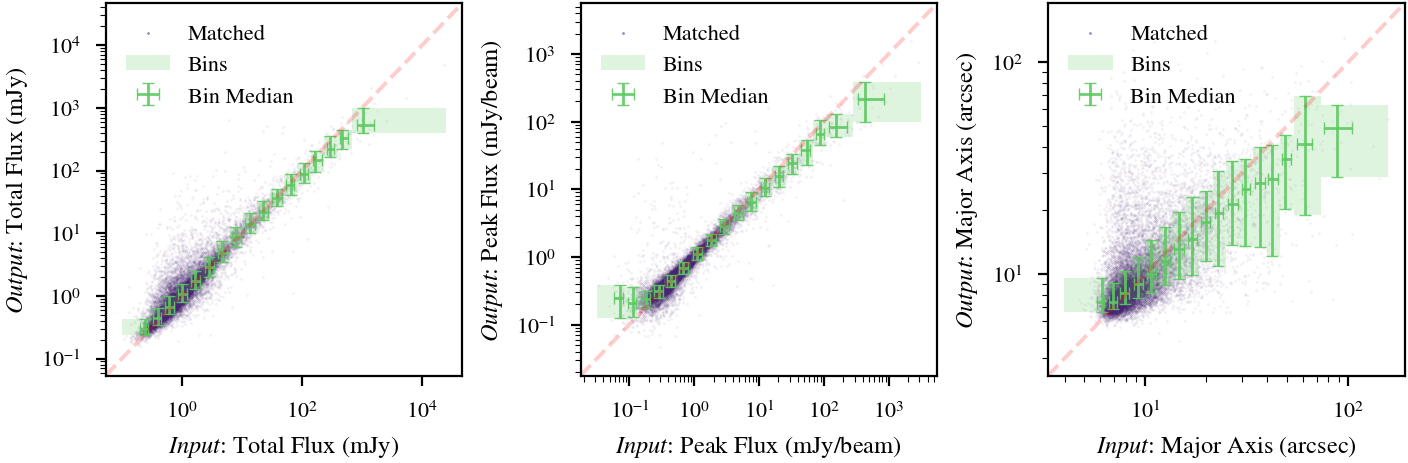}
    \caption{Scatter plots comparing pair-wise input and output values for \textsc{PyBDSF} sources matched between the original images and individual samples produced with the LDM. Error bars indicate the 0.16-0.84 quantiles of the respective bin.}
    \label{fig:ldm_scatter_plots}
\end{figure*}

To further evaluate the sampling quality and degree of control over source properties of the LDM used in isolation, we proceed with the same test cutouts used in Section \ref{sec:results:vq-vae}. We use the catalog context vectors of those test cutouts to sample whole images with the LDM, making no use of inpainting. Hence, we expect the sampled cutouts to contain sources at the same positions, with same brightness and size. However, the sampled images are not expected to be identical to the original cutouts, but rather look like novel realistic images. A selection of samples, together with the original images for visual reference, is shown in Figure \ref{fig:ldm_examples}.\\
To quantify the similarity in source properties, we run \textsc{PyBDSF}, using the same settings employed for the generation of the DR3 source catalog, see \cite{Shimwell+2025} for a detailed description of the procedure. We then match the catalog of detected sources with the input catalogs encoded on the respective catalog context vector, where we use a matching radius of \ang{;;10}. For matched sources, we then compare the three parameters used for sampling, namely total flux, peak flux and major axis. This is shown in Figure \ref{fig:ldm_scatter_plots}. Source matching statistics are shown in Table \ref{tab:ldm_source_matching}. For a single batch of 16 images, on a Nvidia A100 GPU the sampling duration is of \SI{55}{\second}.\\
The examples shown in Figure \ref{fig:ldm_examples} demonstrate the realism of the sampled images, which clearly show a qualitative visual similarity to the original cutouts. Sources appear at the same locations, with similar properties in their brightness and extension. In particular, the spatial correlation seems highly precise, and the relative brightness properties of the sources are well replicated. Also, source extensions are generally well captured for both compact and extended sources, as can be seen in diffuse parts of the bottom source in the third row image. Nonetheless, under close inspection, the chosen examples reveal small variances between the original and sampled cutouts with respect to the controlled properties. For instance, the top right source in the first row image is slightly smaller on the sampled image, and the bottom left source on the fourth row example peaks at a clearly lower maximum flux in the sampled image as compared to the original. The model is also capable of realistically modeling artifacts typically encountered around bright sources, as seen in the bottom row example. This is even captured for sources outside of the FOV, an example of which can be seen in the bottom left corner of the second row example. This capability was explicitly intended for by including neighboring sources in the catalog context array. More examples of bright sources, comparing the original image and the sampled map, are shown in Appendix \ref{app:examples_of_bright_sources}.\\
After matching the \textsc{PyBDSF}-detected sources, counterparts are identified for about \SI{87}{\percent} of input sources. Interestingly, there are more sources detected on the sampled images than were present in the input. A discussion of completeness and purity, and statistics of matched and unmatched sources, are given in Appendix \ref{app:source_matching_statistics}. Figure \ref{fig:ldm_scatter_plots} shows a high degree of correlation over several orders of magnitude between flux values of matched sources, with clear deviations only at the upper and lower limits of the value range, although a relatively small number of individual sources is scattered around the 1:1 line. At increasingly high input fluxes, there is a bias towards lower brightness values in the sampled images, manifesting in both the integrated and peak flux densities. This is consistent with the reconstruction bias of the VQ-VAE observed in Figure \ref{fig:vae_reconstruction_error}. Regarding the the major axis, the correlation is significantly weaker, and matched sources show more variability with respect to their input counterparts, indicating a significantly weaker degree of control over the source sizes as compared to their positions and flux values.

\begin{table}[h]
\caption{Source matching statistics for LDM-Sampled cutouts}    
\label{tab:ldm_source_matching}      
\centering                                      
\begin{tabular}{l l}          
\hline\hline                        
 Group & Source count \\    
\hline                                   
    Total input &  \num{20749}\\
    Total detected & \num{23400}\\
    Matched & \num{18038}\\
\hline                                             
\end{tabular}
\end{table}

\subsection{\textsc{SWIIT}-sampling} \label{sec:results:swiit-sampling}
\begin{figure*}
    \centering
    \includegraphics[width=180mm]{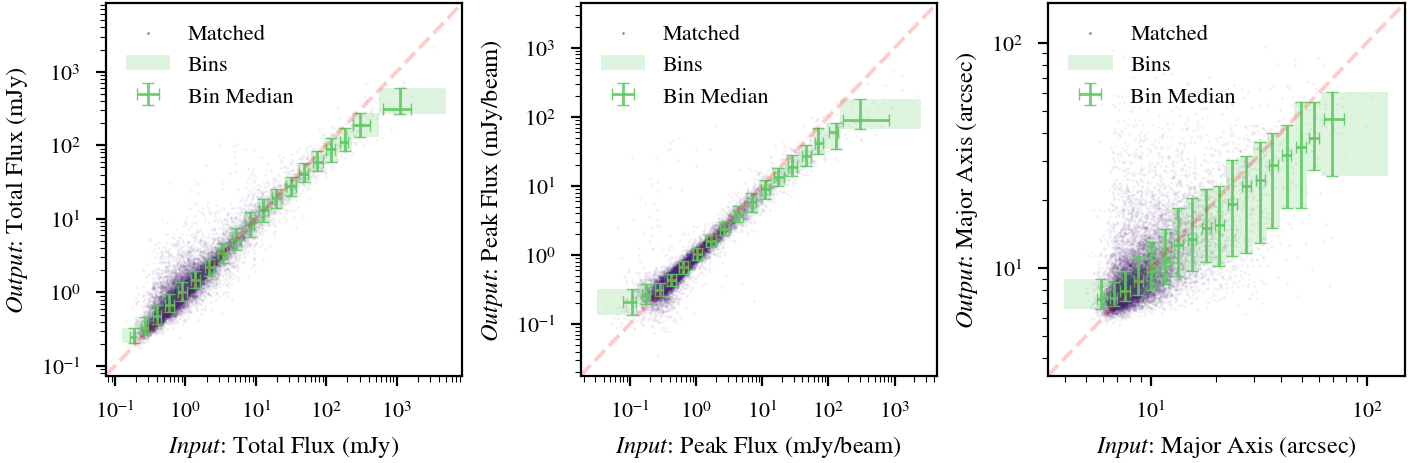}
    \caption{Scatter plots comparing pair-wise input and output values for \textsc{PyBDSF} sources matched between the original cutouts and individual maps produced with the \textsc{SWIIT}-sampler. Error bars indicate the 0.16-0.84 quantiles of the respective bin.}
    \label{fig:swiit_scatter_plots}
\end{figure*}

\begin{figure}
    \centering
    \includegraphics[width=\hsize]{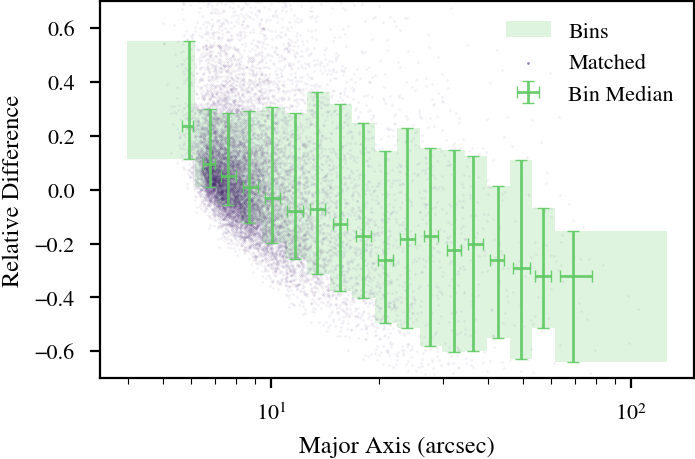}
    \caption{Relative difference between input and output major axis of the matched sources.}
    \label{fig:size_control_relative_difference}
\end{figure}

\begin{figure}
    \centering
    \includegraphics[width=\hsize]{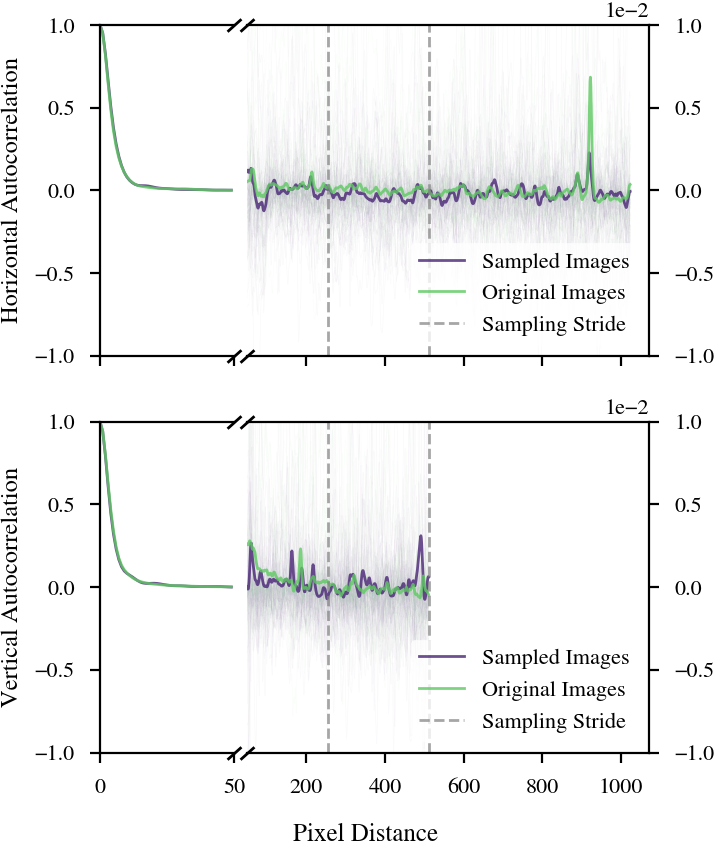}
    \caption{Horizontal and vertical autocorrelation functions of the test cutouts and the corresponding \textsc{SWIIT}-sampled radio maps, calculated as described in Appendix \ref{app:autocorrelation}. Different scalings for both axes are chosen below and above a distance of \SI{50}{px} to adapt visibility in the corresponding ranges.}
    \label{fig:swiit_autocorrelation}
\end{figure}

\begin{figure}
    \centering
    \includegraphics[width=\hsize]{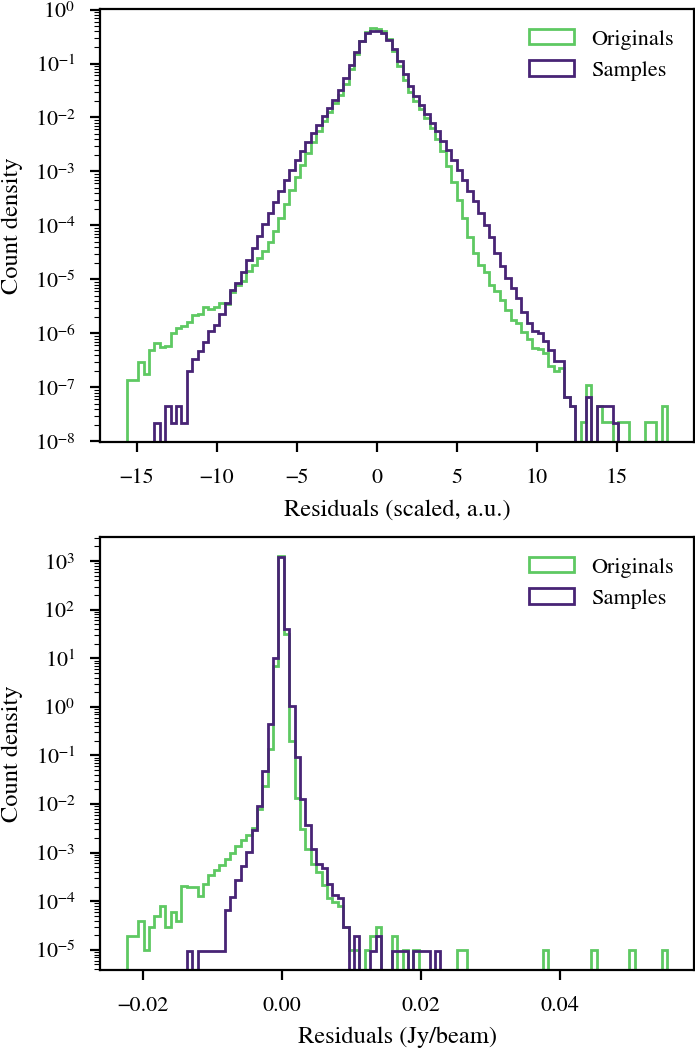}
    \caption{Pixel value distributions of the \textsc{PyBDSF} residual maps of the test cutouts and the corresponding \textsc{SWIIT}-sampled radio maps. The top figure shows the scaled pixel values, the bottom figure shows the unscaled values of the same pixels. }
    \label{fig:residual_pixel_distributions}
\end{figure}

Finally, to evaluate the quality of maps generated with the \textsc{SWIIT}-sampling technique developed for this work, we proceed analogously to the analysis described in Section \ref{sec:results:ldm}. For this, we select 64 LoTSS-DR3 mosaics at random, where from each we extract a central cutout of \num{1024}$\times$\SI{2048}{\px} side length, corresponding to approximately \ang{;25.6;}$\times$\ang{;51.2;} side length or a total FOV of $\SI{260}{\text{arcsec}\squared}$. Sampling 64 of those maps, the total area corresponds to the total area of cutouts sampled in Section \ref{sec:results:ldm}. We also extract the corresponding catalog context map, covering the exact size of the cutout. We then use our \textsc{SWIIT}-sampling procedure to sample maps according to the extracted catalog context, and run the same comparison as in Section \ref{sec:results:ldm}, using \textsc{PyBDSF} and source matching. For better reference, we run \textsc{PyBDSF} on both the sampled maps and original maps, and compare the two resulting catalogs for every map. One example of the sampled map, with its original for comparison, is shown in Figure \ref{fig:swiit_example}. A larger depiction and further examples are shown in Appendix \ref{app:examples}. Source-matching statistics are shown in Table \ref{tab:swiit_source_matching}, parameter scatter plots are shown in Figure \ref{fig:swiit_scatter_plots}. For a batch size of 16 maps sampled in parallel, the total duration for a single map on a Nvidia A100 GPU is \SI{19}{\minute} \SI{1}{\second}.\\
Similar to the individually sampled cutouts, under visual inspection the sampled maps show a high level of realism and control over source properties. In particular, there are no sharp edges or any kinds of artifacts that reveal inconsistencies or discontinuities between individual sampling steps. However, towards the right hand side of Figure \ref{fig:swiit_example}, a repeating stripe-like pattern appears, resembling artifacts associated with bright sources. This is not a pattern typically encountered in the real observations, and is likely an artifact of the sequential sampling. This phenomenon is observed on other examples as well.\\
Comparing the \textsc{PyBDSF} results, \SI{82.4}{\percent} of input sources are matched, slightly less than in the case of individual cutouts although still in the same range. Contrary to the previous case, the total number of sources detected on the sampled maps is slightly below the number of input sources. This is further discussed in Appendix \ref{app:source_matching_statistics}. Comparison between source properties in Figure \ref{fig:swiit_scatter_plots} shows no difference to the previous case, again exhibiting a strong correlation in source brightness with a small bias towards lower values for high flux inputs, and a much weaker correlation in source sizes. In Figure \ref{fig:size_control_relative_difference}, we plot the relative difference between input and output major axis for the matched sources, defined analogously to Equation \eqref{eq:relative_reconstruction_error}. The binned median values clearly decrease with increased input major axis value, possible reasons are discussed in Section \ref{sec:discussion}.\\
To verify the consistency between tile boundaries quantitatively, we calculate the horizontal and vertical normalized autocorrelation function for both the original test cutouts and the sampled maps. An exact description of the calculation is given in Appendix \ref{app:autocorrelation}. The result is shown in Figure \ref{fig:swiit_autocorrelation}. The autocorrelation drops steeply over a distance of few pixels, corresponding to the restoring beam shape. Over pixel distances on the scale of the \SI{256}{px} SWIIT sampling stride, i.e. half the size of an individual image tile, the autocorrelation signal is on the order of \num{e-2}, with occasional peaks caused by bright sources on individual images. Most importantly, no peaks are found at the exact sampling stride or multiples of it, showing no sign of periodic boundary artifacts on the sampled images.\\
To also compare the noise statistics between real and generated radio maps, we plot the pixel distributions of the PyBDSF residual images for both the real test cutouts and generated maps in Figure \ref{fig:residual_pixel_distributions}. We show both the scaled and unscaled pixel values in the top and bottom plot, respectively. The pixel distributions are of the same qualitative shape and cover the same ranges. The distribution for the sampled maps is slightly broader, and the distribution of the original images shows a tail towards the lowest-value pixels that is absent for the sampled images. Small deviations like these are expected, since the model is not trained to replicate the exact noise statistics for any given image, but to produce realistic background signals, which are inherently variable. Nonetheless, the pixel value distributions are well reproduced.\\
Further, to demonstrate the model's capabilities to generalize beyond the sky distributions seen during training, we sample maps from unnatural input catalogs that are not extracted from real LoTSS cutouts. The procedure and resulting images are shown in Appendix \ref{app:unnatural_samples}.\\
Finally, to demonstrate the benefit of using the \textsc{SWIIT} overlap sampling procedure over simple tiling with no context of the neighboring image, we show a comparison in Appendix \ref{app:baseline_comparison} with examples that highlight the possible failure modes of using the latter.\\

\begin{table}[h]
\caption{Source matching statistics for \textsc{SWIIT}-sampled cutouts.}   
\label{tab:swiit_source_matching}      
\centering                                      
\begin{tabular}{l l}          
\hline\hline                        
 Group & Source count \\    
\hline                                   
    Total input & 19805\\
    Total detected & 18325\\
    Matched & 16314\\
\hline                                             
\end{tabular}
\end{table}

\begin{figure*}
    \centering
    \includegraphics[width=180mm]{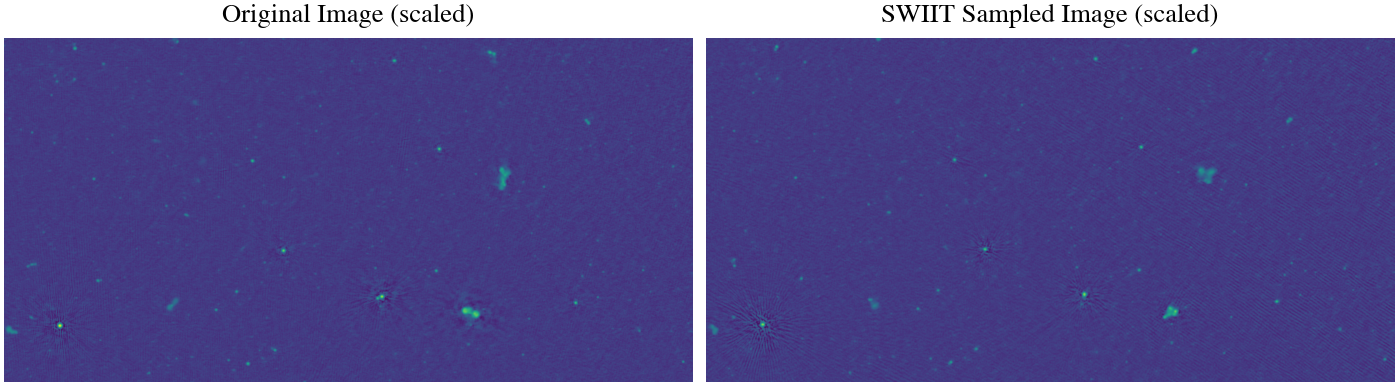}
    \caption{Example of test maps, showing the original map on the left and the \textsc{SWIIT}-sampled map on the right. Note that we do not expect an exact reconstruction, but a realistic image with sources at the same locations, of same brightness and size.}
    \label{fig:swiit_example}
\end{figure*}

\section{Discussion} \label{sec:discussion}
As our provided examples show, the sampled maps replicate highly realistic source morphologies and associated imaging artifacts. Maps generated through \textsc{SWIIT}-sampling have seamless continuity thanks to overlap sampling, blended decoding and the inclusion of neighboring sources in the context vector. The artifacts of repeating patterns observed in several examples are likely an issue of the sequential sampling behavior, where individual steps repeat the pattern beyond a realistic correlation length. This issue could possibly be remedied through further training or dedicated fine-tuning, although it might also be a fundamental weakness of the proposed method. It is possible that such problematic behavior arises for structures whose typical correlation length exceeds the size of sampled images, whereby the model would be trained to continue the structure for most training examples. This is conceivable for stripe-like artifacts typically encountered around very bright sources.\\ 
Our experiments reveal a strong correlation between the flux values of the input context catalog and the sources detected on the sampled images assigned through source matching. This suggests a high degree of sampling control over source location and brightness properties, which is strongly conveyed through visual comparison between original observations and sampled images. The small bias towards lower fluxes for very bright input sources is likely associated to the reconstruction bias of the VQ-VAE. However, the correlation in source sizes between input and detected source sizes is clearly weaker and shows larger scatter. In this work, we parametrize source sizes as the value of the major-axis determined by \textsc{PyBDSF} through image moment analysis of the source model, which is intrinsically reductive, although likely not the limiting factor. Instead, as all input values are encoded through pixel intensity, we hypothesize that it is easy for the network to relate the brightness values in the input vector to the brightness of the source signal at the same location on the output image. In contrast, the relation between pixel intensity on the input vector and source extension on the output image might be less aligned with the model's inductive bias and hence harder to optimize for. This is supported by the fact that the deviation is stronger for larger sources, where corresponding pixels are increasingly further apart from the location of the pixel in the context vector. A possible improvement would be to additionally include the minor axis and positional angle for each of the sources, quantities that are both available in the source catalogs. Those three parameters can then be combined into a single input channel, encoding size information as a binary mask with ellipses shaped according to those size parameters and placed at the respective source positions. This could provide an input signal more closely related to the actual shapes on the training images. Although feasible to implement, this step would require crucial modifications to the training-set preparation, model architecture, and sampling procedure; we therefore leave it for future work.\\
A limitation of the present work is that, since our model is trained on observed radio survey maps, there is no explicit ground-truth sky model against which the generated images can be compared. As a consequence, the method reproduces not only the underlying sky emission but also the observational effects and imaging artifacts present in the training data, which are therefore effectively baked into the simulation rather than being controlled separately. This means that the model is particularly suited for applications in which realistic survey-like images are desired, but less suitable for studies that require a clean separation between astrophysical sky emission and instrument-specific effects. In addition, the present implementation is specific to LOFAR data, and application to surveys from other telescopes would require training on corresponding data from those instruments. The quality of the source catalogs used for training is a limiting factor in this regard, since inaccuracies or incompleteness in the catalog data will directly affect the conditioning information and thus the generated images. Finally, the repeating-pattern artifacts observed in some sequentially generated examples may indicate a limitation of the proposed sampling strategy, although it remains unclear whether this reflects a fundamental property of the method or can be mitigated by further training or fine-tuning.\\
While our experiments show results of remarkable realism and quality, the proposed method works at the expense of high computational cost at both training and inference time. As discussed in Appendix \ref{app:architecture_and_training}, both models require several days of training on a single GPU. In addition, there are two factors that impose a lower limit on the sampling duration. First, by nature of the sequential denoising procedure, diffusion models take several network evaluation steps to sample any individual image. Using classifier-free guidance at 25 sampling time steps, this results in 50 network calls per sampled image. A promising perspective towards alleviating this demand are consistency models \citep{Song+23}, which are trained to reverse the entire forward diffusion process in only a single step. Second, given the sequential nature of the iterative \textsc{SWIIT}-sampling procedure, the number of sampling steps grows proportionally to the area of the sampled map. This latter issue could possibly be mitigated by reducing the overlapping area between subsequent sampling steps, which is currently half the width of an individual image. This would reduce the proportionality factor between area of the sampled image and \textsc{SWIIT}-sampling steps, although it might happen at the expense of decreased consistency between individual samples generated at different steps. Another option would be to parallelize the sampling of independent steps, trading sampling speed for increased GPU space per map.\\

\section{Summary and conclusions}
In this work, we implemented a generative model for synthesizing realistic individual radio survey map images of \SI{512}{px} side length, a size that is to our knowledge unprecedented for generative models in the context of astronomical image generation. Further, we developed \textsc{SWIIT}-sampling, a novel technique for sampling radio sky survey maps of arbitrary size through iterative use of a LDM trained for image inpainting. By designing a customized function for scaling the pixel values of training images, the large dynamic range of radio maps can be dealt with in a consistent way over the entire dataset, rather than by scaling every image individually. This makes radio observational data suitable for computer vision tasks while still conserving universal flux information. By engineering a context vector that encodes spatial information of radio sources present on the images, as well as their brightness and size properties, we facilitate effective control over sources present on the sampled images. By also including information of sources in the immediate vicinity outside of the FOV, we account for a correct modeling of spillover artifacts and thereby allow for coherent iterative sampling in a sliding-window approach. This makes it possible to sample realistic images at arbitrary sizes, circumventing current size limitations of deep learning models for image generation.\\
Large synthetic radio maps generated by our model can be used for developing and testing different kinds of methods applied to radio astronomical image data. For instance, in the development of automated source detection and identification software, controlled sampling allows for the targeted generation of test images with specific desired properties, covering edge cases like e.g. very extended sources or highly clustered fields, where algorithms might require dedicated testing. In a similar way, our model can be used for generative data augmentation to improve un- or self-supervised learning of source classification models \citep[e.g.][]{BaronPerez+2025}.\\
While this work focuses on radio sky maps, the method can be employed for generating any kinds of large-scale mock image data, given the availability of enough training images. For instance, \cite{Mishra+2025} introduce a similar approach for the prediction of HI intensity cubes from input halo mass density cubes, using a method they call Latent Overlap Sampling. They train a DM on three-dimensional HI data, conditioned on halo mass density fields, and derive predictions by iteratively sampling larger cubes in a way analogous to our \textsc{SWIIT}-sampling. Our work differs in a few crucial aspects: we use two-dimensional image data rather than three-dimensional cubes, we employ a LDM rather than a pure DM, and we explicitly account for the inpainting capabilities both in the model input and training methodology.\\
Our proposed procedure can also be used as an emulator for otherwise more computationally expensive modeling steps. For instance, in \cite{VicanekMartinez+2025}, synthetic survey maps are produced by first generating raw visibility data from a synthetic sky model, and then obtaining the synthetic radio image through data reduction and deconvolution. This last, highly computationally expensive step could be replaced by training an LDM conditioned on the sky model instead. Although this would require the previous generation of a large dataset of synthetic model-image pairs, this procedure could be implemented with the simulated responses of different radio telescopes, making this an interesting application for mock observations used in the development of next-generation instruments like the Square Kilometre Array \citep{Dewdney+2009}.

\section{Data availability}
The code implemented and used for this work is publicly available under \url{https://github.com/tmartinezML/glori}.

\begin{acknowledgements}
We thank the referee for a constructive report. MB and TVM acknowledge support by the Deutsche Forschungsgemeinschaft under Germany’s Excellence Strategy – EXC 2121 Quantum Universe – 390833306 and via the KISS consortium (05D23GU4) funded by the German Federal
Ministry of Education and Research BMBF in the ErUM-Data action plan.\\
\end{acknowledgements}

\bibliographystyle{aa}
\bibliography{references}

\begin{appendix}

\section{Pixel value scaling} \label{app:pixel_scaling}

\begin{figure}[h!]
    \centering
    \includegraphics[width=\hsize]{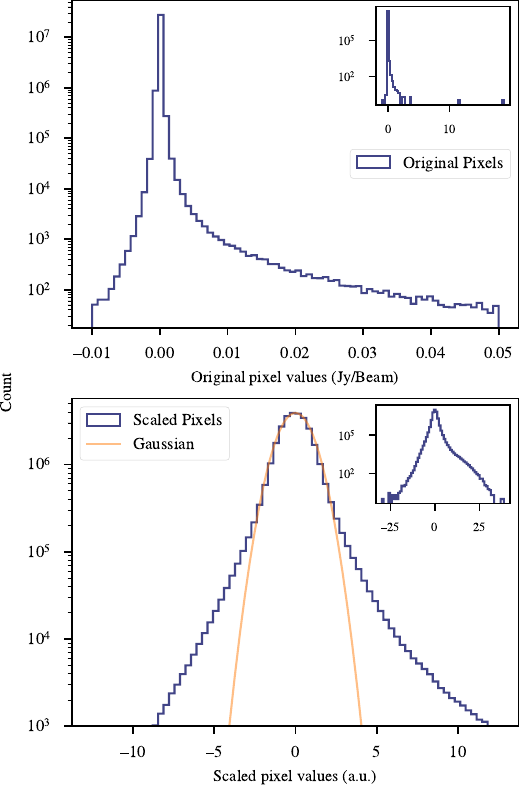}
    \caption{Distributions of original (top) and scaled (bottom) pixel values. Inset plots show the entire range, while the large plots show a selected range of high counts.}
    \label{fig:pixel_value_distirbutions}
\end{figure}

The pixel values of the LoTSS mosaics follow a highly skewed and asymmetric distribution that spans several orders of magnitude, as shown in the top plot of Figure \ref{fig:pixel_value_distirbutions}. To prepare the input images for efficient model training, we apply a scaling designed to make the pixel value distribution approximately have zero-mean and unit variance. To develop the scaling, we extract a random set of \num{10000} pixels for each one of the \num{817} mosaics of the second LoTSS data release, which are used to make the shown histograms. The function $\gamma$ employed to obtain the scaled pixel values from the original values $x$ is defined as 
\begin{equation}
    \gamma_\pm(x) \coloneqq a_\pm \cdot \ln{\left( \frac{s}{a_\pm} \cdot x + 1\right)}, \label{eq:pixel_scaling}
\end{equation}
where the index $_\pm$ indicates that different values are used depending on the sign of $x$, i.e. $a_+$ if $x \geq 0$ and $a_-$ if $x < 0$. The parameters $a_\pm$ and $s$ are manually tuned, such that the distribution of scaled values is approximately zero-mean and unit variance, yet still unimodal. The resulting values are shown in Table \ref{tab:scaling_function_values} and the distribution of scaled pixel values is shown in the bottom plot of Figure \ref{fig:pixel_value_distirbutions}. To invert the scaling, we employ the inverse function 
\begin{equation}
    \gamma^{-1}_\pm(y) = \frac{a_\pm}{s} \cdot \left( e^{ y/a_\pm} - 1\right). \label{eq:pixel_scaling_inverse}
\end{equation}
This scaling is applied only to pixel values. The values of the conditioning vectors in units of \si{\milli Jy}, \si{\milli Jy \per beam} and \si{arcsec} remain unscaled.

\begin{figure*}[h!]
    \centering
    \sidecaption
    \includegraphics[width=12cm]{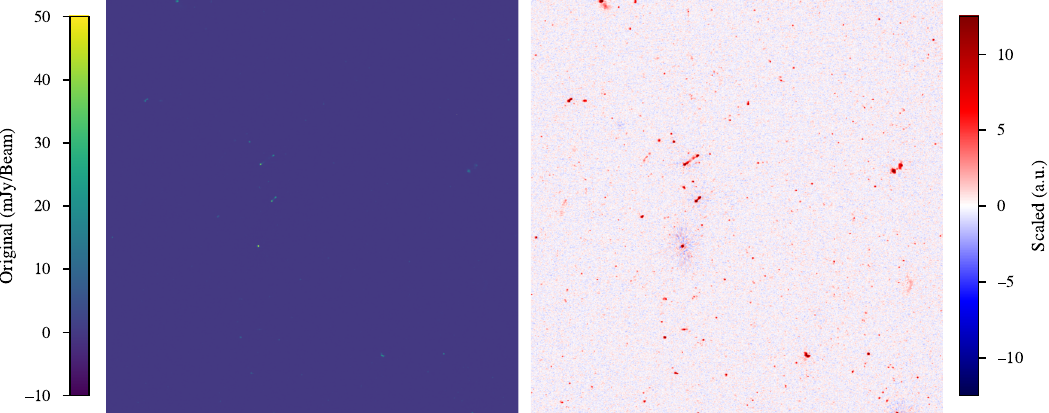}
    \label{fig:pixel_scaling_example}
    \caption{Example of original (left) and scaled (right) image cutout.}
\end{figure*}

\begin{table}[h!]
    \caption{Values used for the scaling function $\gamma$ as defined in Equation \eqref{eq:pixel_scaling}.}              
    \label{tab:scaling_function_values}      
    \centering                                      
    \begin{tabular}{l l}          
        \hline\hline                        
        Parameter & Value \\    
        \hline                                   
            $a_+$ & \num{0.813} \\
            $a_-$ &  \num{0.858}\\
            $s$ & 2500 \\
        \hline                                             
    \end{tabular}
\end{table}

\FloatBarrier

\section{Optimal tiling pattern for cutout extraction}\label{app:mosaic_tiling}

The optimal pattern for dividing any LoTSS-DR3 mosaic into square-shaped cutouts of a side length $L$ is determined as follows. We first calculate the maximum valid mosaic width $W$ as the space between the left-most and right-most valid pixel of the image. We then divide this patch horizontally into $h$ bins of width $L$, whereby $h \coloneqq \left\lfloor{\frac{W}{L}}\right\rfloor$. These bins are arranged symmetrically around the center of $W$. For each of these columns, we individually calculate the maximum valid height $H$ between the lower and upper bound of valid pixels. The lower bound is the highest pixel of all the bottom valid pixels for each individual pixel column within the respective bin. Likewise, the upper bound is the lowest of all the top valid pixels for each pixel column. We then divide the column vertically into $v$ squares of side length $L$, whereby $v \coloneqq \left\lfloor{\frac{H}{L}}\right\rfloor$. Again, the squares are stacked symmetrically around the center of $H$. Examples for a typical mosaic of circular shape are shown in Figure \ref{fig:cutout_tiling}.

\FloatBarrier
\section{Model architecture and training details}\label{app:architecture_and_training}
For every model that we train, we calculate validation loss at regular intervals. We keep parameter checkpoints of the 5 best performing models, measured by lowest validation loss over a fixed number of validation batches, and after training we use the best-performing checkpoint parameters for evaluation.
\subsection{VQ-VAE}
A list of architecture choices is given in Table \ref{tab:VQ-VAE_Architecture}, in reference to the architecture explained in \ref{sec:model-architecture-and-training} and illustrated in Figure \ref{fig:VQ-VAE_architecture}. Training hyperparameters including the weights for the different loss terms are shown in Table \ref{tab:VQ-VAE_Hyperparams}. We use a codebook of size 2048 with three channels in latent space. The discriminator network training, together with the corresponding generator loss term, is only introduced after \num{10000} iterations. We adopt several improved training techniques from \cite{Huh+2023}. These include the affine reparametrization of codebook vectors with learnable shared mean and variance for which the learning rate is scaled by a factor 10, the synchronized update rule with $\nu = 0.2$, and the initialization of code vectors with K-means clustering. Code vectors that are not used for 20 iterations are re-initialized dynamically during training. We train the model for \num{4e5} iterations with a generator loss weight of \num{0.5}, and further train it for \num{e5} more iterations setting the generator loss weight to \num{0.1}. We observe that this additional training improves reconstruction without introducing blurriness. This leads to the interpretation that the value of $\alpha_\mathrm{Gen}$ can possibly be further optimized for. The base training run is performed on a single Nvidia A100 GPU and runs for \SI{7}{\day}~\SI{5}{\hour}~\SI{17}{\minute}~\SI{31}{\second}. The additional training run is performed on a Nvidia H100 GPU and runs for \SI{1}{\day}~\SI{7}{\hour}~\SI{15}{\minute}~\SI{24}{\second}\\

\begin{table}[h!]
    \caption{Details of the implemented VQ-VAE Encoder-Decoder architecture.}
    \label{tab:VQ-VAE_Architecture}      
    \centering
    \begin{tabular}{l l}
        \hline\hline  
        Parameter & Value \\    
        \hline                        
        Resolution levels & $(256, 128, 64)$ \\
        Initial channels & \num{128} \\
        Channel multipliers & $(1, 2, 4) $ \\
        Norm. groups & \num{32} \\
        Attention & Bottleneck layer only \\
        Attention heads & \num{2} \\
        Attention head channels & \num{32} \\
        Parameters & \num{4.5e7}\\
        \hline
    \end{tabular}
    \tablefoot{
         Resolution levels indicate the image sizes in pixels on the different levels in the Encoder-Decoder architecture. The number of channels at different levels is given as a product of the initial channels and corresponding channel multiplier.
    }
\end{table}

\begin{table}[h!]
    \caption{Training hyperparameters for the VQ-VAE.}
    \label{tab:VQ-VAE_Hyperparams}      
    \centering
    \begin{tabular}{l l}
        \hline\hline  
        Parameter & Value \\    
        \hline                        
        $\alpha_\mathrm{Rec}$ & 1 \\
        $\alpha_\mathrm{Rec,unsc}$ & 100 \\
        $\alpha_\mathrm{Gen}$ & \num{0.5} (\num{0.1}) \\
        $\alpha_\mathrm{Commit}$ & \num{1} \\
        $\beta$ & \num{0.9} \\
        Iterations & \num{4e5} (\num{1e5}) \\
        Batch size & \num{10} \\
        Learning rate (VQ-VAE) & \num{5e-6} \\
        Learning rate (Discriminator) & \num{2e-5} \\
        \hline
    \end{tabular}
    \tablefoot{Parameters in brackets indicate values used for the additional training run.}
\end{table}

\FloatBarrier
\subsection{DM}\label{app:architecture_and_training:DM}
A list of architecture choices for the denoiser U-Net is given in Table \ref{tab:Unet-architecture}. Choices for the architecture of the smaller U-Net used as a catalog context encoder are given in Table \ref{tab:Context-encoder-Unet-architecture}. For the latter, we use gated attention \citep{Qiu+2025} in the decoder module between the CNN block outputs and the skip connections from the encoder module. This is done to improve the learning of global context across the entire area of the catalog encoding map. Hyperparameters used for training the denoiser are given in Table \ref{tab:Denoiser_hyperparams}. We use dropout layers at the output of each CNN block. Noise levels $\sigma_\mathrm{Train}$ during training are sampled from a log-normal distribution, such that $\ln \sigma_\mathrm{Train} \sim \mathcal{N}(P_\mathrm{Mean}, P_\mathrm{Std})$, corresponding values are shown in Table \ref{tab:Denoiser_hyperparams}. We use the Heun 2nd order solver \citep{Ascher1998ComputerMF} for sampling with noise schedule
\begin{equation}
    \sigma_{1 \leq \mathrm{t} \leq N - 2}=
        \left(\sigma_{\max }{ }^{\frac{1}{\rho}}+\frac{t}{N-1}\left(\sigma_{\min }{ }^{\frac{1}{\rho}}-\sigma_{\max}{}^{\frac{1}{\rho}}\right)\right)^\rho,
\end{equation}
where $\sigma_0 = \sigma_{\max}$, $\sigma_{N-1} = \sigma_{\min}$ and $\sigma_N = 0$. We use values $\sigma_{\max}=80$, $\sigma_{\min}=\num{2e-3}$ and $\rho = 7$. All images in this work are sampled with 25 timesteps.\\
We apply a scaling to the neural network as described in \cite{Karras+22}, which reduces variations in the magnitudes of signals and gradients. For a neural network $F_\theta$, the denoiser $D_\theta$ is evaluated as 
\begin{equation}
    D_\theta(\mathbf{x}, \sigma) = c_\mathrm{skip}(\sigma) \cdot \mathbf{x} + c_\mathrm{out}(\sigma)\cdot F_\theta\left(c_\mathrm{in}(\sigma)\cdot\mathbf{x},c_\mathrm{noise}(\sigma)\right),
\end{equation}
with constants defined as
\begin{align}
    &c_\mathrm{skip}(\sigma) = \sigma_\mathrm{data}^2 \cdot \left(\sigma^2 + \sigma_\mathrm{data}^2\right)^{-1},\\
    &c_\mathrm{out}(\sigma) = \sigma \cdot \sigma_\mathrm{data} \cdot \left(\sigma^2 + \sigma_\mathrm{data}^2\right)^{-\frac{1}{2}}, \label{eq:c_out}\\
    &c_\mathrm{in}(\sigma) = \left(\sigma^2 + \sigma_\mathrm{data}^2\right)^{-\frac{1}{2}},\\
    &c_\mathrm{noise}(\sigma) = \frac{1}{4}\ln \sigma,
\end{align}
whereby $\sigma_\mathrm{data} = 0.5$ is a fixed value. The parameter $c_\mathrm{out}$ defined in Equation \eqref{eq:c_out} is the same one used for the training loss in Equation \eqref{eq:denoiser_loss}.\\
During training, inpainting masks are varied randomly in the following way. We sample with uniform probabilities between four possible masks that correspond to sampling the entire image, sampling the right half, sampling the bottom half, and sampling the bottom right corner. Depending on which of the inpainting masks is used, we then blank parts of the catalog context array, meaning we set all values in the blanked part to zero. This is done in order to replicate sampling at the edge of a sky map, where extended context information is not available. In case of the inpainting mask covering the entire image or right half, the top quarter of the catalog context array is blanked with a probability of 0.9. Similarly, in case of the inpainting mask covering the entire image or bottom half, the left quarter of the catalog context array is blanked with a probability of 0.9. Further, if the top quarter is not blanked, we blank the bottom quarter with a probability of 0.1, and if the left quarter is not blanked, we blank the right quarter with a probability of 0.1.\\ 
We first train the model with the given batch size for \num{2e5} iterations. We then employ a second stage of training where, to have more accurate gradient updates, we effectively double the batch size by using gradient accumulation. This second training stage runs again for \num{2e5} iterations. Both training runs are run on a single Nvidia A100 GPU and take a total time of \SI{10}{\day}~\SI{2}{\hour}~\SI{16}{\minute}~\SI{6}{\second}.\\

\begin{table}
    \caption{Details of the implemented DM denoiser U-Net architecture.}
    \label{tab:Unet-architecture}      
    \centering
    \begin{tabular}{l l}
        \hline\hline  
        Parameter & Value \\    
        \hline                        
        Resolution levels & 3\\
        Initial channels & 256 \\
        Channel multipliers & (1, 2, 4)\\
        Norm. groups & \num{32} \\
        Attention & (No, Yes, Yes) \\
        Attention heads & \num{4} \\
        Attention head channels & \num{32} \\
        Parameters & \num{2.6e8}\\
        \hline
    \end{tabular}
    \tablefoot{
         Resolution levels indicate the image sizes in pixels on the different levels in the U-Net architecture. The number of channels at different levels is given as a product of the initial channels and corresponding channel multiplier.
    }
\end{table}

\begin{table}
    \caption{Details of the catalog context encoder network, which is itself a U-Net.}
    \label{tab:Context-encoder-Unet-architecture}      
    \centering
    \begin{tabular}{l l}
        \hline\hline  
        Parameter & Value \\    
        \hline                        
        Resolution levels & 3 \\
        Initial channels & 64 \\
        Channel multipliers & (1, 2, 2)\\
        Norm. groups & \num{4} \\
        Attention & (No, Yes, Yes) \\
        Attention heads & \num{8} \\
        Attention head channels & \num{8} \\
        Parameters & \num{6.7e6}\\
        \hline
    \end{tabular}
    \tablefoot{
         Resolution levels indicate the image sizes in pixels on the different levels in the U-Net architecture. The number of channels at different levels is given as a product of the initial channels and corresponding channel multiplier.
    }
\end{table}

\begin{table}
    \caption{Training hyperparameters for the DM denoiser.}
    \label{tab:Denoiser_hyperparams}      
    \centering
    \begin{tabular}{l l}
        \hline\hline  
        Parameter & Value \\    
        \hline                        
        Iterations & \num{2e5} (\num{2e5}) \\
        Batch size & \num{24} (\num{48}) \\
        Learning rate & \num{5e-6} \\
        $P_\mathrm{Mean}$ & \num{-2.5} \\
        $P_\mathrm{Std}$ & \num{1.8} \\
        Dropout rate & \num{0.1} \\
        Context dropout rate & \num{0.1} \\
        \hline
    \end{tabular}
    \tablefoot{Values in brackets indicate the parameters used for the second training stage.}
\end{table}

\FloatBarrier

\section{Examples of bright sources}\label{app:examples_of_bright_sources}
\begin{figure*}
    \centering
    \includegraphics[width=180mm]{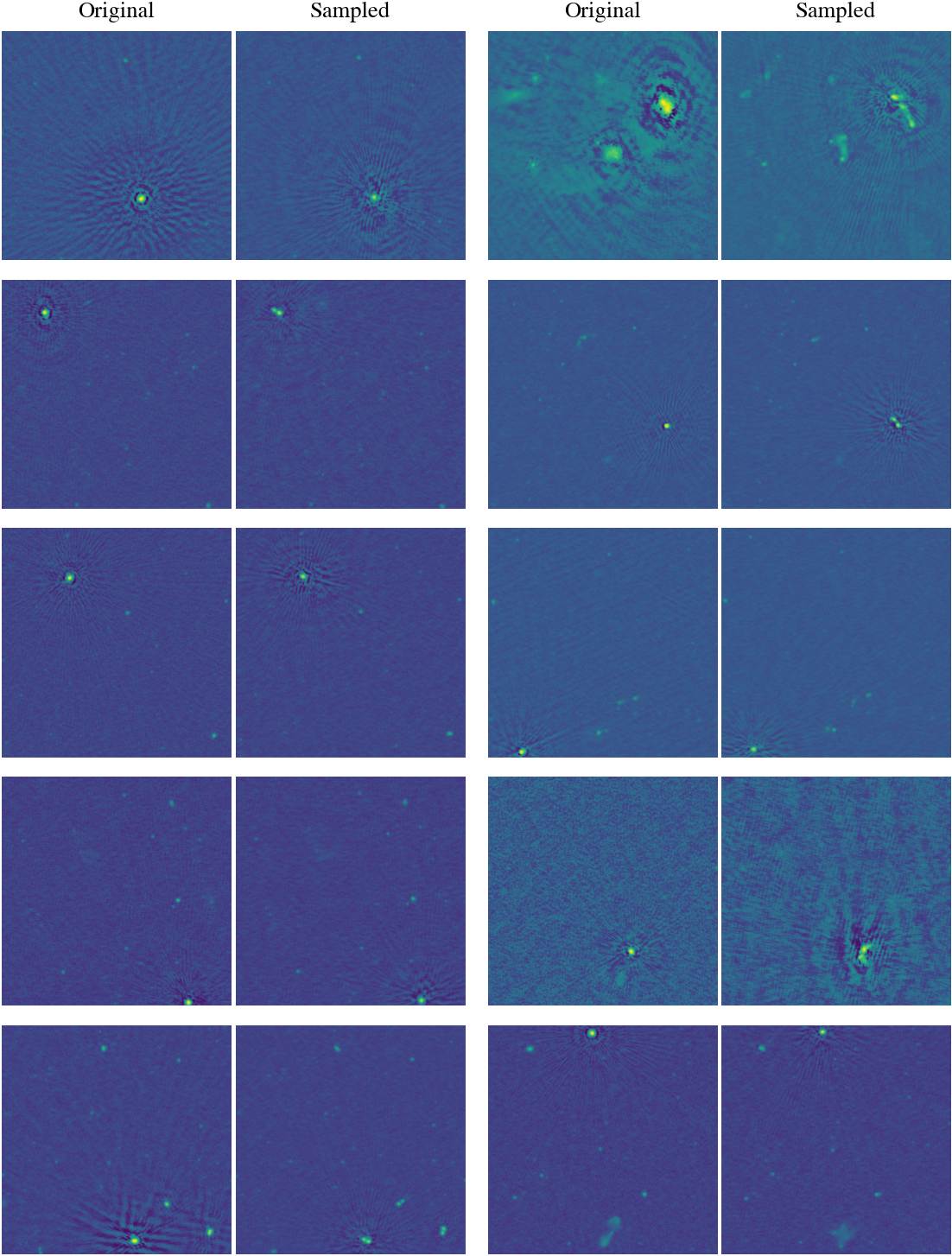}
    \caption{Example pairs of test cutouts with \SI{512}{px} (\ang{;12.8;}) side length containing bright sources, showing the original on the left and the corresponding LDM-sampled image on the right.}
    \label{fig:LDM_examples_bright_sources}
\end{figure*}

In Figure \ref{fig:LDM_examples_bright_sources}, we show a selection of test cutouts and corresponding LDM-sampled images that contain bright sources. These are typically surrounded by ring- and spoke-like sidelobe artifacts that arise from incomplete removal of the interferometer's point-spread function. Since residual sidelobes also depend on local noise, source blending, and direction-dependent variations, equally bright sources can show different artifact patterns. We therefore expect the LDM to qualitatively generate those artifacts, but no exact quantitative replication.

\FloatBarrier

\onecolumn

\section{Source matching statistics}\label{app:source_matching_statistics}
\begin{figure*}[h!]
    \centering
    \sidecaption
    \includegraphics[width=12cm]{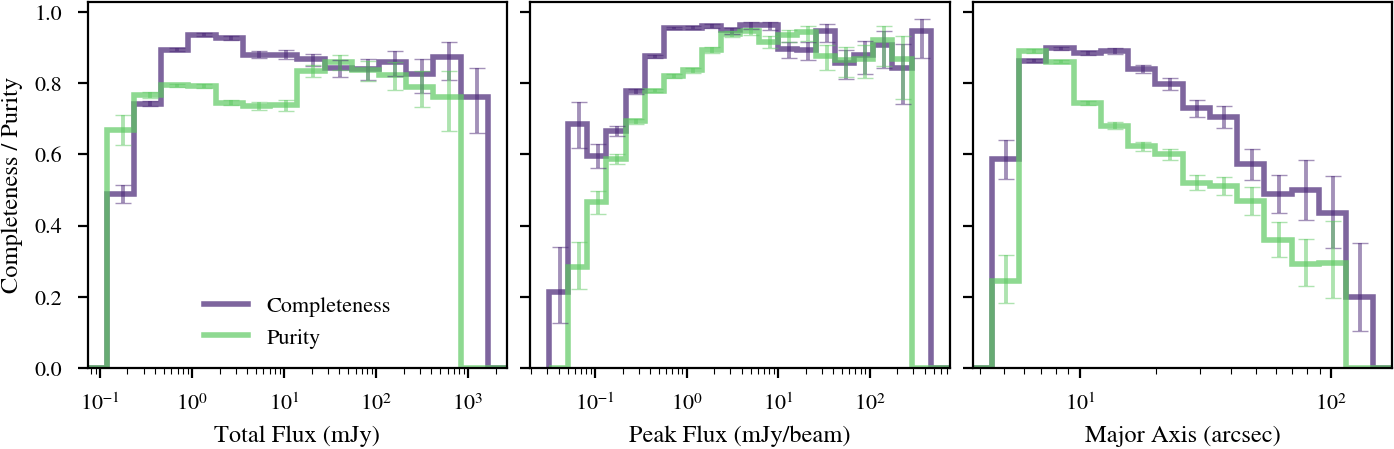}
    \caption{Completeness and purity plots for the LDM-Sampled test cutouts, binned over the different source properties.}
    \label{fig:completeness_purity_LDM}
\end{figure*}

\begin{figure*}[h!]
    \centering
    \sidecaption
    \includegraphics[width=12cm]{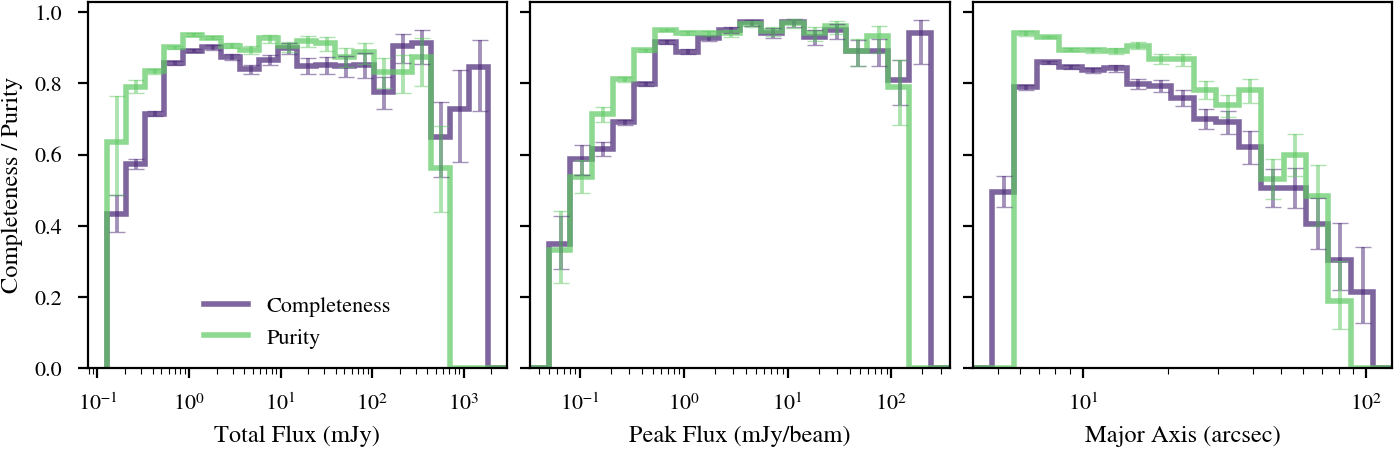}
    \caption{Completeness and purity plots for the \textsc{SWIIT}-Sampled test maps, binned over the different source properties.}
    \label{fig:completeness_purity_SWIIT}
\end{figure*}

\begin{figure*}[h!]
    \centering
    \sidecaption
    \includegraphics[width=12cm]{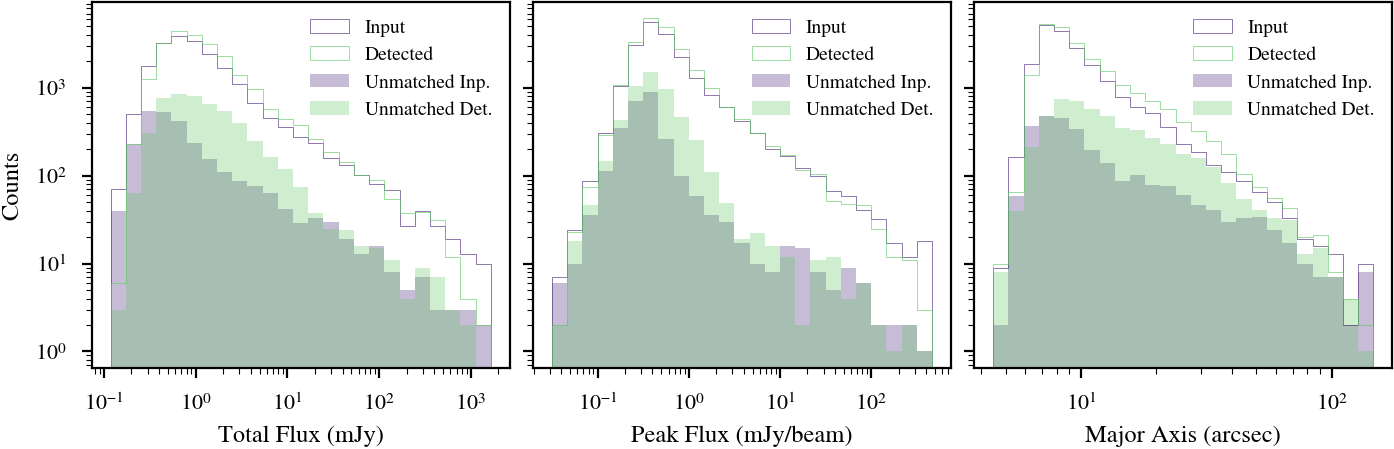}
    \caption{Histograms over the different source properties of all input and detected sources for the LDM-Sampled test cutouts, showing also the statistics of the unmatched sub-populations.}
    \label{fig:overdetections_histograms}
\end{figure*}

For the LDM-sampled images, we detect a larger number of sources than is present in the input catalog. This is likely related to the \textsc{PyBDSF} noise rms estimation used to determine the detection threshold, which becomes less reliable when estimated on smaller images. As a result, more low-flux sources are detected, since lowering the detection threshold adds more sources than are removed when it is increased, owing to the asymmetric flux distribution of sources. This is indicated by the middle panels of Figures \ref{fig:completeness_purity_LDM} and \ref{fig:completeness_purity_SWIIT}, which show the completeness and purity of matched sources as a function of different source properties for the LDM- and \textsc{SWIIT}-sampled images, respectively. For the LDM-sampled images, the completeness of matched sources is higher at lower peak flux values, but the purity is correspondingly lower than for the \textsc{SWIIT}-sampled maps. In addition, most of the over-detections occur in the low-flux regime, as shown in Figure \ref{fig:overdetections_histograms}, which presents the distributions of the total number of sources and the unmatched sources across different source properties. This is also supported by visual inspection. However, interpretation is not straightforward, because nearest-neighbor matching may lead to mismatches, for example when bright emission regions around bright sources in one image are matched to the bright source in the other image, or when different components of the same source are detected as separate sources. The fact that fewer sources are detected in the \textsc{SWIIT}-sampled images than in the input catalog may likewise be related to the detection threshold, since the largest reduction in completeness occurs in the low-flux regime. It is possible that the detection threshold is increased in some images due to the sampling artifacts described in Section \ref{sec:results:swiit-sampling}.

\FloatBarrier
\twocolumn

\section{Calculation of vertical and horizontal autocorrelation}\label{app:autocorrelation}

The horizontal and vertical autocorrelation of an image $\rho_\mathrm{h}(k)$ and $\rho_\mathrm{v}(k)$ as a function of pixel distance $k$ are calculated analogously. We first normalize the image $\mathbf{x}$ by subtracting the mean pixel value, giving $\tilde{\mathbf{x}} \coloneqq \mathbf{x} - \bar{\mathbf{x}}$. We then proceed to calculate the autocovariance for every row (column) $m$ as 
\begin{equation}
    C_m(k) = \sum_{n=0}^{L-1-k} \tilde{x}_{m,n}\,\tilde{x}_{m,n+k}, \qquad k = 0,\dots,L-1,
\end{equation}
where $L$ is the image width (height), and indices $i,j$ indicate the position of a pixel $\tilde{x}_{i,j}$. Since only $L-k$ pixel pairs contribute at distance $k$, we normalize by the corresponding pair count, and the total autocorrelation profile $C(k)$ is obtained by averaging over all rows (columns), giving
\begin{equation}
    C(k) = \frac{1}{L}\sum_{m=0}^{L-1} \frac{C_m(k)}{L-k}.
\end{equation}
Finally, we normalize by the zero-distance value to obtain a coefficient bound in $[-1, 1]$, resulting in 
\begin{equation}
    \rho(k) = \frac{C(k)}{C(0)}.
\end{equation}

\FloatBarrier

\section{Sampled images with unnatural source distributions}\label{app:unnatural_samples}

\begin{figure*}
    \centering
    \includegraphics[width=180mm]{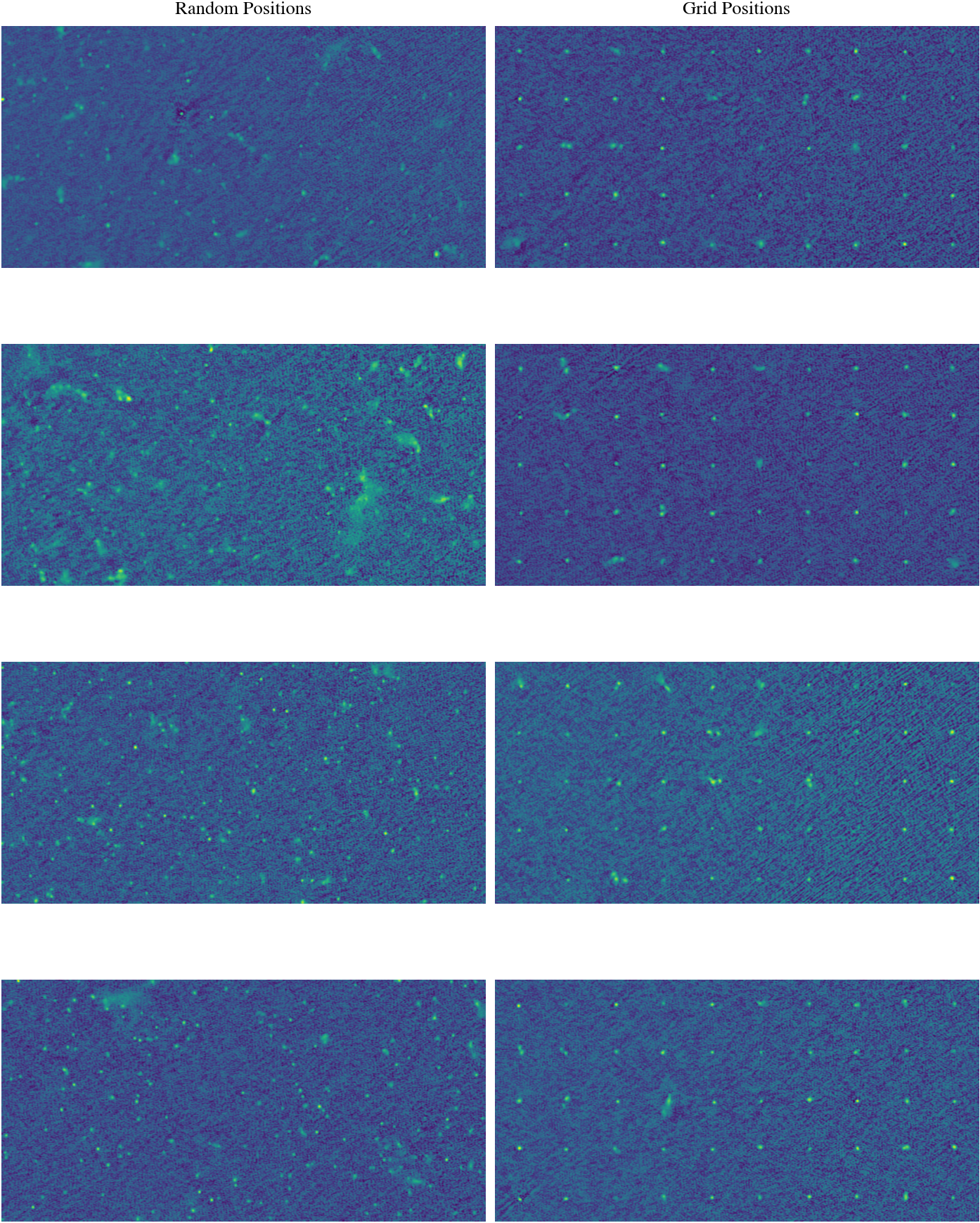}
    \caption{Examples of sampled images with unnatural source distributions.}
    \label{fig:swiit_examples_unnatural}
\end{figure*}

To demonstrate the capabilities of our model to sample images beyond the distributions seen during training, we produce samples with unnatural source distributions in the following way.
To obtain realistic source properties, we produce a three-dimensional histogram with logarithmic binning in the space of integrated flux density, peak flux density and major axis. We sample corresponding values for individual sources by first choosing a bin with probability proportional to its counts, and then selecting values between the bin edges with uniform probability in logarithmic space. We repeat this procedure for number of sources desired for the respective image.\\
We proceed to sample images of \num{512}$\times$\SI{1024}{\px} side length, corresponding to \ang{;12.8;}$\times$\ang{;25.6;}, in two different versions. For the first version, we choose a number of 200 sources, which corresponds to the upper limit of count density found in the test cutouts. We distribute the source positions randomly with uniform probability. For the second version, we position 50 sources per image on a regular grid. Examples of the resulting images are shown in Figure \ref{fig:swiit_examples_unnatural}.\\
While these examples show the expected results, referring to them as realistic in this context would strictly speaking be ill-defined, and the concept of image quality becomes ambiguous. We also note that image quality breaks down when the limits are pushed further, e.g. by heavily over-crowding the field. However, these cases are unlikely to be relevant for any real scientific application.

\FloatBarrier
\onecolumn
\section{Comparison of \textsc{SWIIT}-sampling and simple tiling}\label{app:baseline_comparison}
\begin{figure*}[h!]
    \centering
    \sidecaption
    \includegraphics[width=12cm]{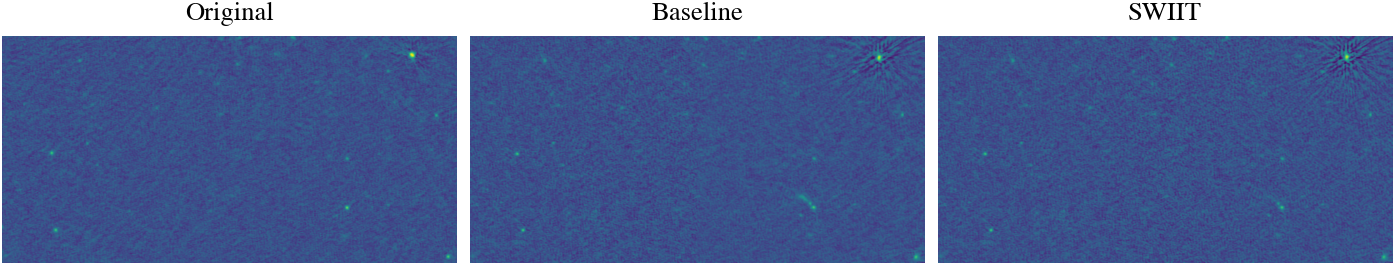}
    \caption{Comparison of original map, map sampled through simple tiling ("baseline"), and map sampled with our \textsc{SWIIT}-sampling procedure. The baseline sample shows inconsistent background properties.}
    \label{fig:baseline_comparison_inconsistent_bg}
\end{figure*}

\begin{figure*}[h!]
    \centering
    \sidecaption
    \includegraphics[width=12cm]{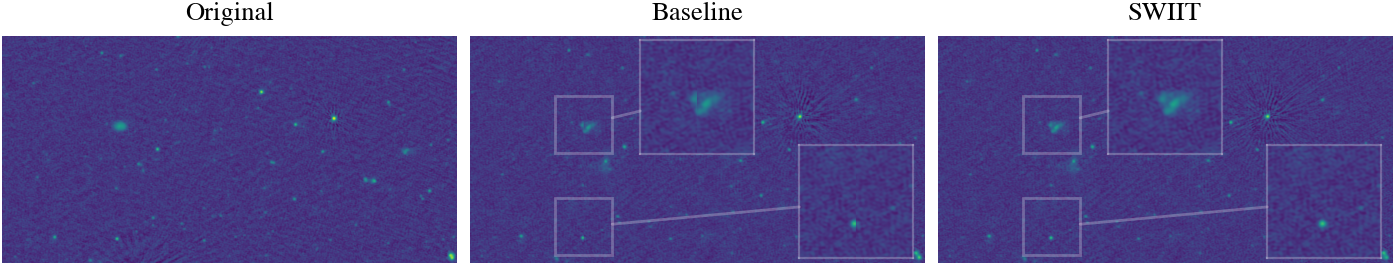}
    \caption{Comparison of original map, map sampled through simple tiling ("baseline"), and map sampled with our \textsc{SWIIT}-sampling procedure. The baseline sample shows edge discontinuities.}
    \label{fig:baseline_comparison_edge_disc}
\end{figure*}

To demonstrate the benefits of using our \textsc{SWIIT}-sampling procedure over simple tiling of individual images with no overlap context, we provide a visual comparison of an example map sampled in both ways. For this, we choose an example from the test cutouts and extract a smaller patch of size \num{128}$\times$\SI{256}{\px}, corresponding to \ang{;3.2;}$\times$\ang{;6.4;}. This size is sampled within three steps of our \textsc{SWIIT} procedure, serving as a minimal example. We then sample an image in two different ways, where in both cases we use the context vector from the catalog of the extracted patch, and the same initial seed noise for both images. The first version samples every step individually, using a stride of half the image length like the \textsc{SWIIT} procedure, but without inpainting context and simply placing individual samples on the final map. We refer to this version as the baseline. The second version uses the \textsc{SWIIT}-sampling procedure as described. Figures \ref{fig:baseline_comparison_inconsistent_bg} and \ref{fig:baseline_comparison_edge_disc} show two different examples of the described procedure, i.e. the comparison for two different example patches, each highlighting a different failure mode of the simple tiling procedure.\\
The first example in Figure \ref{fig:baseline_comparison_inconsistent_bg} shows how simple tiling can lead to inconsistent background characteristics across individual images. The baseline version shows a clear, visible distinction between the left and right half of the map, where each looks individually realistic, but the coherence over the entire map is broken. This issue does not appear for the \textsc{SWIIT}-sampled version.\\
The second example in Figure \ref{fig:baseline_comparison_edge_disc} shows how simple tiling can cause artifacts and discontinuities when sources fall on the boundaries between different sampling steps. As highlighted by the magnified sources in the inset plots, there are clear edges that produce unnatural source morphologies, which again do not appear on the \textsc{SWIIT}-sampled maps.

\FloatBarrier
\twocolumn

\section{Further examples of sampled images}\label{app:examples}
\begin{figure}[h!]
    \centering
    \includegraphics[width=\hsize]{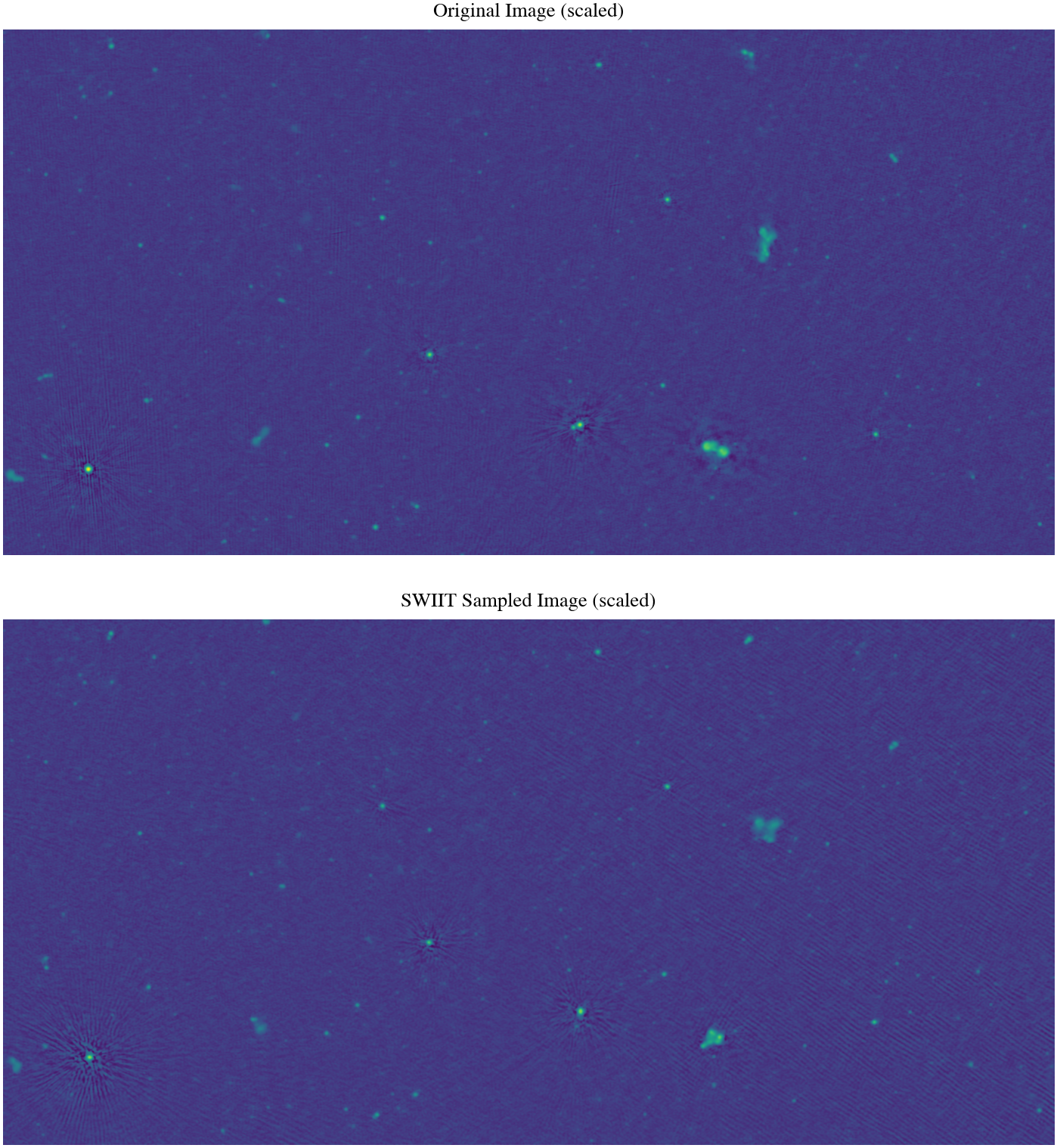}
    \caption{Example of test maps, showing the original map on top and the \textsc{SWIIT}-sampled map on the bottom. This is the same example shown in Figure \ref{fig:swiit_example}. Note that we do not expect an exact reconstruction, but a realistic image with sources at the same locations, of same brightness and size.}
    \label{fig:swiit_example_appendix}
\end{figure}

\begin{figure}[h!]
    \centering
    \includegraphics[width=\hsize]{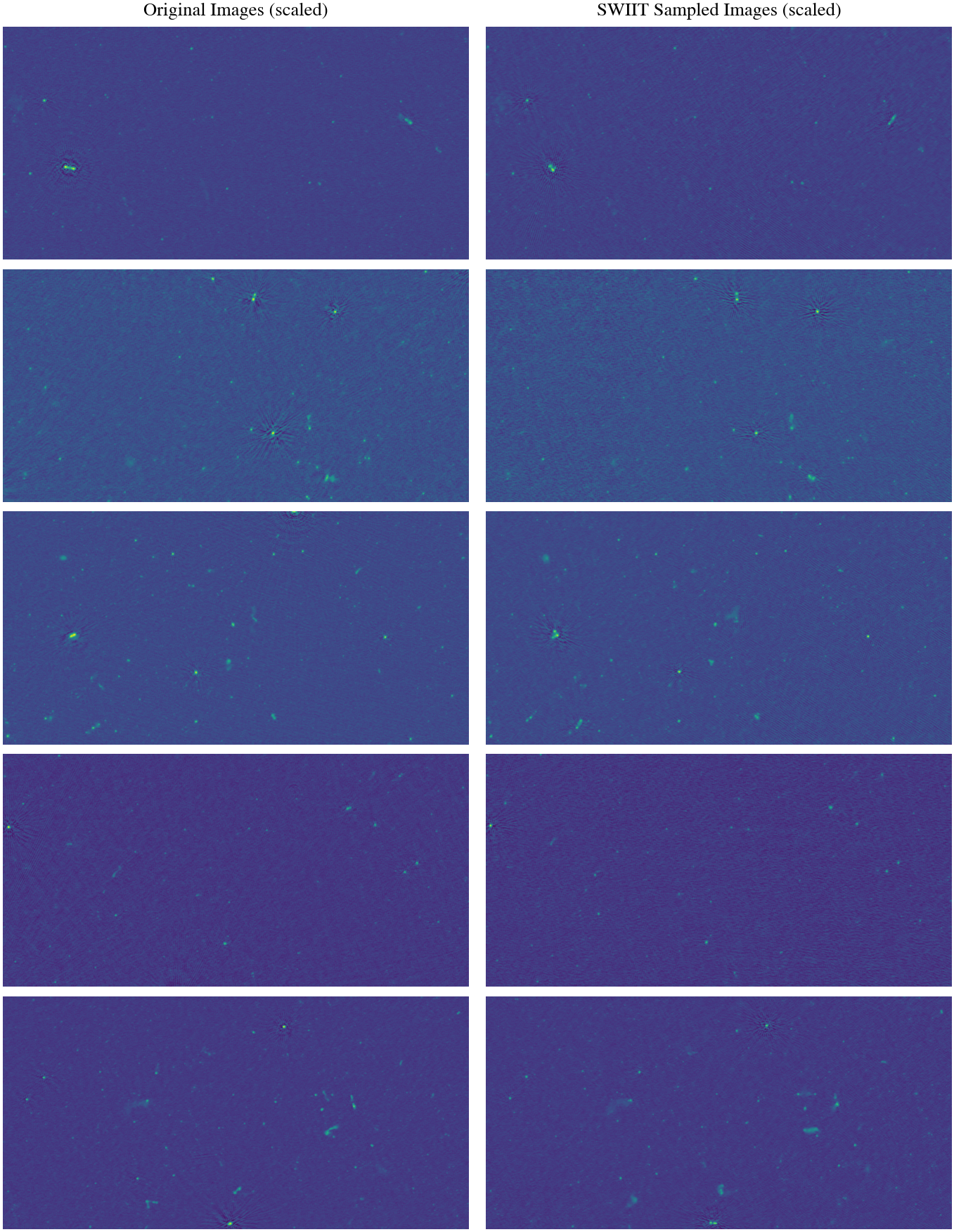}
    \caption{Example pairs of test maps, showing the original map on the left and the \textsc{SWIIT}-sampled map on the right. Note that we do not expect an exact reconstruction, but a realistic image with sources at the same locations, of same brightness and size.}
    \label{fig:swiit_examples_appendix}
\end{figure}

\begin{figure*}[h!]
    \centering
    \includegraphics[width=180mm]{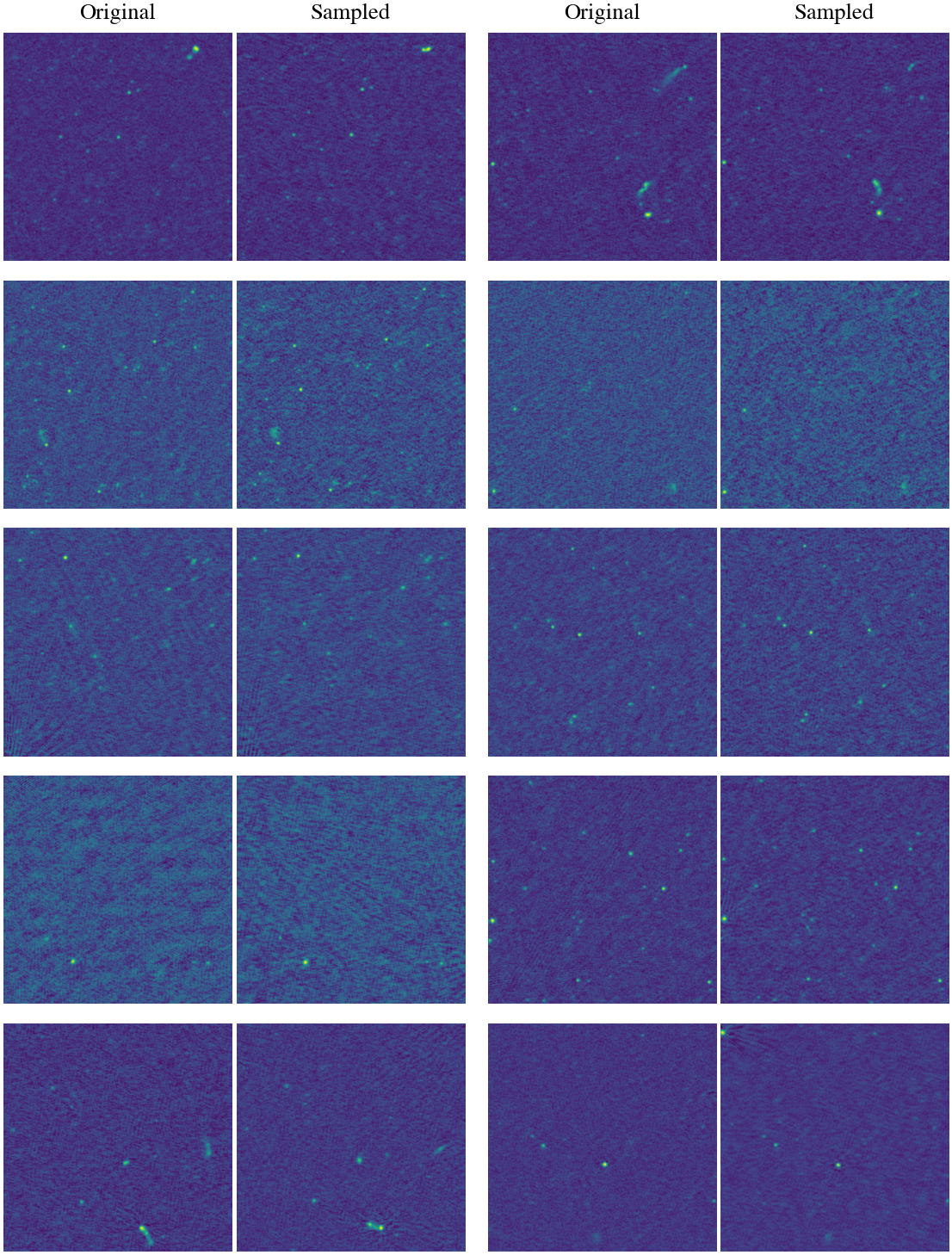}
    \caption{Example pairs of test cutouts, showing the original cutout on the left and a generated sample on the right. Note that we do not expect an exact reconstruction, but a realistic image with sources at the same locations, of same brightness and size.}
    \label{fig:ldm_examples_appendix}
\end{figure*}

In Figure \ref{fig:ldm_examples_appendix}, we show an additional selection of individual test cutouts with LDM samples based on the corresponding catalog context array. In Figure \ref{fig:swiit_example_appendix}, we show the a larger view of the same \textsc{SWIIT}-sampled map shown in Figure \ref{fig:swiit_example}, and we show further examples of \textsc{SWIIT}-sampled maps in Figure \ref{fig:swiit_examples_appendix}.

\end{appendix}

\end{document}